\documentclass[superscriptaddress,aps,preprint,nofootinbib,showpacs]{revtex4}
\usepackage{graphicx,color,amsmath, epsfig, rotating}
\usepackage{feynarts}
\usepackage{float}

\newcommand{\bae}{\begin{eqnarray}}
\newcommand{\eae}{\end{eqnarray}}

\newcommand{\beq}{\begin{equation}}
\newcommand{\eeq}{\end{equation}}
\newcommand{\beqa}{\begin{eqnarray}}
\newcommand{\eeqa}{\end{eqnarray}}
\newcommand{\ba}{\begin{array}}
\newcommand{\ea}{\end{array}}
\newcommand{\no}{\nonumber}
\newcommand{\bea}{\begin{eqnarray}}
\newcommand{\eea}{\end{eqnarray}}
\def\delr            {\!\stackrel{\leftrightarrow}{\partial^\mu}\!}

\newcommand{\meee}{{\mu \to \bar{e}ee}}

\newcommand{\meg}{{\mu \to e\gamma}}
\newcommand{\teg}{{\tau \to e\gamma}}
\newcommand{\tmg}{{\tau \to \mu\gamma}}

\begin{document}
\title{\boldmath Lepton Flavour Violating $\tau$ and $\mu$ decays \\
induced by scalar leptoquark}

\author{ Rachid Benbrik}

\affiliation{LPHEA, Department of Physics, FSSM, Cadi Ayyad University,
P.O.B. 2390, Marrakech 40 000, Morocco.}

\affiliation{
Department of Physics, National Cheng-Kung University, Tainan 701, Taiwan
}

\affiliation{
National Center for Theoretical Sciences, Hsinchu 300, Taiwan.}

\author{Mohamed Chabab}

\affiliation{LPHEA, Department of Physics, FSSM, Cadi Ayyad University,
P.O.B. 2390, Marrakech 40 000, Morocco.}

\author{Gaber Faisel}

\affiliation{
Egyptian Center for Theoretical Physics, Modern University
for Information and Technology , Faculty of Engineering, AlHadaba
AlWusta,  AlMokattam, Cairo, Egypt.
}
\affiliation{
Faculty of education, Thamar University, Yemen.
}

\date{\today}


\begin{abstract}
We show that  the scalar leptoquark Yukawa couplings
generate a significant lepton flavour violation. We compute the
light scalar leptoquark contributions to the branching ratios (Br)
of the lepton flavour violating (LFV) decays $\ell \to \ell_i
\ell_j \bar{\ell}_j$ and $\ell \rightarrow \ell^{\prime} \gamma$ with
(i,j = e,$\mu$).  We discuss the role of the relevant input
parameters to these decay rates which are the Yukawa couplings
($h_{a\ell}$) with ($a=u,c,t$), the light scalar mass $M_{S_{1}}$ 
and the mixing angle $\sin2\theta_{LQ}$. 
We investigate the experimental limits
from $(g-2)_\mu$, $\mu-e$ conversion and $\pi \to e \nu_e, \mu \nu_\mu$ 
to get constraint on the input parameter space. 
We predict that the upper limits on the branching ratios of
$\tau \to \ell_i \ell_j \bar{\ell}_j $ can reach the experimental current
limits. We also show  that it is possible to accommodate
both $\tau \to \ell_i \ell_j \bar{\ell}_j$ and $\tau \to \ell\gamma$ branching ratios
for certain choices of LQ parameters.

\end{abstract}
\maketitle

%=======================================================
\section{Introduction}

Lepton-flavor violation (LFV), 
if observed in a future experiment, is an evidence of new
physics beyond the standard model, 
because the lepton-flavor number is conserved in
the standard model. 
Since the processes 
%do not suffer from a large ambiguity due to
%the hadronic matrix elements, 
%detailed analysis of the LFV processes will reveal some
%properties of the high-energy physics.
%They
are theoretically free from the non perturbative hadronic effects
they provide accurate predictions for the decay rates and the
branching ratios (Br) of these processes. Furthermore, they are
theoretically rich as they  carry considerable information about
the free parameters of the used model. On the other hand, the
experimental work which has been done regarding these decays
motivates their theoretical studies.  For instance, experimental
prospect for $\mu \to e \gamma$ is promising with the recent
commencement of the MEG experiment which will probe Br$(\mu \to e
\gamma) \approx 10^{-13}$  two orders of magnitude beyond the
current limit. B factories search for the decay mode $\tau \to
\ell_i \ell_j \bar{\ell_j}$ at the  $e^+ e^-$ experiment with
upper limits in the range Br$(\tau \to\ell_i \ell_j
\bar{\ell_j})\le (2-8)\times 10^{-8}$~\cite{Aubert:2003pc}.
Searches for $\tau \to \mu\mu\bar{\mu}$ can be performed at the
Large Hadron Collider (LHC) where $\tau$ leptons are copiously
produced from the decays of $W$, $Z$, $B$ and $D$, with anticipated
sensitivities to Br$(\tau \to \mu\mu\bar{\mu}) \approx
10^{-8}$~\cite{Giffels:2008ar}. The decay $\mu \to ee\bar{e}$ of
which there is a strict bound Br$(\mu \to ee\bar{e})\le 10^{-12}$
is a strong constraint on the parameter
space~\cite{Bellgardt:1987du}.

The present experimental upper limits for the branching ratios  of
$\ell \to \ell_i \ell_j \bar{\ell}_j$ and $\ell \to \ell^{\prime}\gamma$ decays are given
by~\cite{Aubert:2003pc,Bellgardt:1987du} \bae \label{llld} {\rm
Br}(\tau \to \ell_i \ell_j \bar{\ell}_j) \sim  10^{-8},\qquad {\rm
Br}(\meee) \sim   10^{-12},
%{\rm Br}(\tmmm) & < & 3.2 \times 10^{-8},\\
%{\rm Br}(\teee) & < & 3.6 \times 10^{-8},\\
%{\rm Br}(\tmme) & < & 3.7 \times 10^{-8},\\
%{\rm Br}(\tmee) & < & 2.0 \times 10^{-8},\\
%{\rm Br}(\teme) & < & 2.7 \times 10^{-8},\\
%{\rm Br}(\temm) & < & 2.3 \times 10^{-8},
\eae and \cite{Aubert:2005ye, Aubert:2005wa,Brooks:1999pu}
\bae
\label{lllm} {\rm Br}(\tmg) & < & 6.8 \times 10^{-8},\\ \no
{\rm Br}(\teg) & < & 1.1 \times 10^{-7},\\ \no
{\rm Br}(\meg) & < & 1.2
\times 10^{-11}. \eae
Within the SM, the Brs of LFV decays are extremely
small. On the other hand, the difference  between the experimental
value of the  muon anomalous magnetic moment $a_\mu = (g-2)/2$ and
its  SM prediction  is given
by\cite{Bennett:2006fi,Yao:2006px,Miller:2007kk}
\begin{eqnarray}
\label{gm2} \Delta a_\mu = a^{\rm exp}_\mu - a^{\rm SM}_\mu =
(29.5 \pm 8.8)\times 10^{-10},
\end{eqnarray}
with a discrepancy of 3.4 $\sigma$.  In spite of the substantial
progress in both experimental and  theoretical sides, the
situation is not completely clear yet. However, the possibility
that the present discrepancy may arise from the errors in the
determination of the hadronic leading-order contribution to
$\Delta a_\mu$ seems to be unlikely as  argued  in
Ref.~\cite{Passera:2008jk}. There are many
attempts, in the literature,  to explain this discrepancy through considering new
physics beyond SM~\cite{Bigi:1985jq,Czarnecki:2001pv,Ellis:2007fu}.

One of the possibilities  for physics beyond the Standard Model is
the four-color symmetry between quarks and leptons introduced by Pati-Salam~\cite{Pati:1974yy}. The prediction of the
existence of gauge leptoquarks, which are rather heavy according
to the current available data, is a direct consequence of this
symmetry.

The current bounds on the leptoquarks production are set by
Tevatron, LEP and HERA~\cite{Wang:2004cj}. Tevatron experiments
have set limits on the scalar leptoquarks masses $M_{LQ} > $ 242
GeV. On the other hand, the  limits that have been set by LEP and
HERA experiments  are model dependent. The search for these novel
particles will be continued at the CERN LHC. Preliminary
studies at the LHC experiments,  ATLAS~\cite{Mitsou:2004hm} and CMS~\cite{Abdullin:1999im}, indicate that clear signals can be
observed for masses  up to 1.2 TeV.

  Our aim in this paper is to analyze the branching ratios  for all
processes given in Eqs.(\ref{llld})- (\ref{lllm}) in the context
of the LQ model. These LFV processes are generated at loop level
through exchanging  scalar LQ particles which transmit the lepton
flavour mixing from the Yukawa couplings to the observed charged
lepton sector. Previous studies of such decays were performed
extensively by theorists~\cite{Okada:1999zk}. In the present study of
these decay channels, the light scalar leptoquark effects
to $\ell \to \ell_i \ell_j \bar{\ell}_j$ are discussed in detail, 
namely the contributions of the
photon and Z boson penguins and box diagrams. 
Also, we include the predictions for $\ell \to \ell_i \ell_i \bar{\ell}_i$ 
channels correlated with $\ell \to \ell^{\prime} \gamma$ rates which are
interesting within the framework we use. Furthermore, we take into
account $(g-2)_\mu$, $\mu-e$ conversion and $\pi \to e \nu_e, \mu \nu_\mu$ 
constraints imposed on the input parameter space. 
This is carried here by considering the parametrization introduced in~\cite{Benbrik:2008si} for the case of the $\ell \to \ell_i \ell_j
\bar{\ell}_j$ decays.\\
The paper is organized as follows: In  Section \ref{formalism}, we
 list the relevant terms  of the scalar  leptoquark Lagrangian to the LFV
 decays and  the analytical expressions of the
 scalar leptoquark contributions to $a_\mu$
 and $\ell \to \ell^{\prime} \gamma$
 decays. The analytical results of the LFV decays $\ell \to
\ell_i \ell_j \bar{\ell}_j$  will be presented in Sec.III.
In Sec.IV, we derive the constraints that can be imposed on some
leptoquark Yukawa couplings obtained using $\mu - e $ conversion.
The numerical results for  $\tau$ and $\mu$ decays will be
presented in Sec.V. Finally, Sec.VI will be devoted to the
conclusion.

%--------------------------------------------
\section{Leptoquark Basics}
\label{formalism}
\subsection{Scalar Leptoquark Interactions}
%-------------------------------------------

In this section we list the relevant terms of the scalar
leptoquark Lagrangian to our LFV decay modes. We consider
isosinglet scalar leptoquarks. The effective Lagrangian that describes
the leptoquark interactions in the mass basis can be written as~\cite{CHH1999,Lagr1}:
\begin{eqnarray}
% LQ
 {\mathcal{L}}_{LQ}
 &=&
 \label{lag}
 \overline{u^c_a} \bigg(h^{'}_{ai} \Gamma_{k,S_R} P_L+h_{ai}\Gamma_{k,S_L} P_R \bigg) e_i S^*_k
 +\overline{e_{j}} \bigg( h^{'*}_{aj} \Gamma^\dagger_{S_R,k} P_R+h^*_{aj} \Gamma^\dagger_{S_L,k} P_L
 \bigg) u^c_a S_k
 \\\no
 &-& e Q_{(u^c)} A_\mu \overline{u^c_a} \gamma^\mu u^c_a - ieQ_{S} A_\mu S^*_{k} \delr S_{k}
 + ieQ_{S} \tan\theta_W Z_\mu S^*_{k} \delr S_{k}
 \\\no
 &-& \frac{e }{s_W c_W} Z_\mu \overline{u^c_a}\gamma^\mu
 \bigg( (T_{3(u^c)} - Q_{(u^c)} s^2_W) P_R - Q_{(u^c)} s^2_W P_L \bigg)
 u^c_a,
\end{eqnarray}
where $k = 1,2$ are the leptoquark indices, $T_3 = -1/2$, $Q_{u^c}
= -2/3$ are quark's isospin and electric charge respectively, $Q_S
= -1/3 $ is the electric charge of the scalar leptoquarks $S_k$,
$a$ is up-type quark flavor indices, $i,j$ are lepton flavor
indices, $c_{W} = \cos\theta_W$ and $s_{W} = \sin\theta_W$.
The $\Gamma_{k, S_{L(R)}}$ are elements of leptoquark mixing matrix
that bring $S_{L(R)}$ to the mass eigenstate basis $S_k$:
%---------
\begin{eqnarray}
S_L = \Gamma^{\dagger}_{S_L, k} S_k, \qquad S^*_R = \Gamma_{k,S_R}
S^*_k,
\end{eqnarray}
Here $S_{L(R)}$ denotes the field  associated with the
$\overline{e_{j}} P_{L(R)} u^c_a$ terms in
${\mathcal{L}}_{LQ}$~\cite{CHH1999}. Note that in the no-mixing
case ($\Gamma=1$), $S_{1(2)}$ reduce to $S_{L(R)}$ which are called
chiral leptoquarks as they only couple to quarks and leptons in
certain chirality structures.
Finally, the couplings $h$ and $h^{\prime}$ are 3 by 3 matrices that give
rise to various LFV processes and must be subjected to the
experimental constraints. In this work we do not intend to explore
the effects of all possible leptoquark interactions.
Instead, we try to demonstrate that a simple scalar leptoquark
model can provide rich and interesting LFV phenomena.
%========================================
\subsection{Muon anomalous magnetic moment $(g-2)_{\mu}$}
%========================================

%**************************************************************
%****              Figure 2
%*******************************************************
%**************************************************************
%****              Figure 2
%*******************************************************
\begin{figure}
\begin{center}
%\FIGURE{
\vspace{2cm}
\input{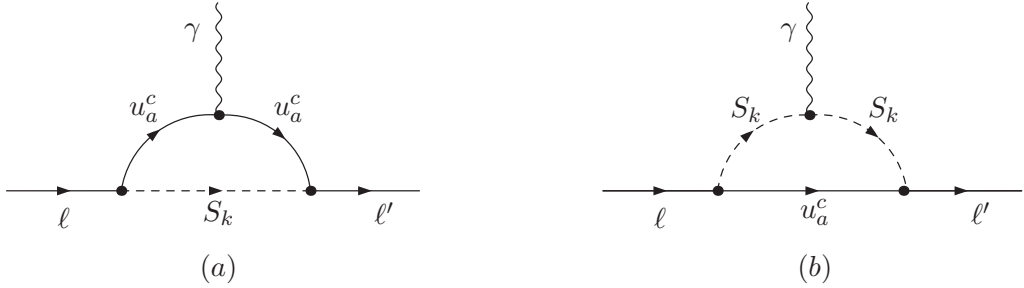}
\vspace{-14.2cm} \caption{Feynman diagrams contributing to $\ell
\to \ell^\prime \gamma$,
 ${S}_k$ denotes the scalar leptoquark with $k=1,2$ and $u^c_a$ denotes up-type quark   with  $a=1,2,3$. }
\label{fig:gm2-diagrams} 
%}
\end{center}
\end{figure}
%===============================================

%===============================================
The LQ interaction can generate muon anomalous magnetic moment and
resolve the discrepancy between theoretical and experimental
results. The corresponding one-loop diagrams are shown in
Fig.~\ref{fig:gm2-diagrams}(a)- \ref{fig:gm2-diagrams}(b) 
where $\ell=\ell^\prime = \mu$.
The extra contribution to $a_\mu $ arising from the LQ model due to
quark and scalar leptoquark one-loop contribution is given by
\begin{eqnarray}
a^{LQ}_{\mu} &=&\no - \frac{N_c m^2_\mu}{8 \pi^2} \sum_{a=1}^{3}
\sum_{k=1}^{2}
 \frac{1}{M^2_{S_k}} \bigg[ \big(|h_{a\mu} \Gamma_{k,S_L}|^2  + |h'_{a\mu} \Gamma_{k,S_R}|^2 \big)
\big( Q_{(u^c)} F_{2}(x_{ka})- Q_{S} F_{1}(x_{ka}) \big) \\&& -
  \frac{m_{(u^c_a)}}{m_\mu}  {\rm Re} \big(h'_{a\mu} h^*_{a\mu} \Gamma^{+}_{S_R,k}\Gamma_{k,S_L} \big)
\big(Q_{(u^c)} F_{3}(x_{ka}) - Q_{S} F_{4}(x_{ka}) \big) \bigg],
\label{eq:a_LQ}
\end{eqnarray}
In the above expression, $N_c = 3$, $Q_{S} = -1/3$, $Q_{u^c} =
-2/3$. The kinematic loop functions $F_{i}$ $(i=1,...,4)$ depend
on the variable $x_{ka} = m^2_{(u^c_a)} / M^2_{S_k}$, their
expressions are given in the appendix B.

Clearly, the use of leptoquark contribution to saturate the deviation
shown in Eq.(\ref{gm2}) leads to constraint leptoquark masses 
$M_{S_{k}}$ (k=1,2), mixing angle
$\theta_{LQ}$ and the Yukawa couplings ($h_{a\mu}$, $h^{(\prime)}_{a\mu}$).

%======================================
\subsection{$\ell \to \ell^{\prime} \gamma$}
%=====================================

In this subsection, we give the  expression for the amplitude of
$\ell \to \ell'\gamma$ which is generated by exchange of scalar leptoquark. According to the gauge invariance, the amplitude can be written
as:
\begin{eqnarray}
\label{ampge} i{\mathcal{M}}^{\gamma}&=&ie\bar{u}(p_2) \bigg(
F^{\gamma}_{2RL} P_{L} + F^{\gamma}_{2LR} P_R\bigg)
(i\sigma_{\mu\nu}q^\nu)
 u(p_1)\varepsilon^{\mu *}_{\gamma},
\end{eqnarray}
where $\varepsilon_\gamma$ is the polarization vector and $q= p_1
- p_2$ is the momentum transfer. For the amplitude of leptoquark
exchange at one-loop level, as depicted in
Fig.~\ref{fig:gm2-diagrams} with $\ell \neq \ell^{\prime}$, we have
\begin{eqnarray}
\label{lepq} F^{\gamma}_{2LR} &=&\no \frac{N_c}{16
\pi^2}\sum_{a=1}^{3} \sum_{k=1}^{2}
  \frac{1}{M^2_{S_k}}\Bigg[
\big(m_\ell h'_{a\ell} h^{'*}_{a\ell'} \Gamma^{\dagger}_{S_R,k}
\Gamma_{k,S_R}    + m_{l'} h_{a\ell} h^{*}_{a\ell'}
\Gamma^{\dagger}_{S_L,k} \Gamma_{k,S_L}  \big)
\\\no &&\times
\big(Q_{(u^c)}F_{2}(x_{ka})
  - Q_{S}F_{1}(x_{ka})\big)
\\ &&- m_{(u^c_a)}
\big(h_{a\ell} h^{'*}_{a\ell'} \Gamma^{\dagger}_{S_R,k}
\Gamma_{k,S_L}\big) \big(Q_{(u^c)}F_{3}(x_{ka}) -  Q_{S}
  F_4 (x_{ka})\big)\Bigg],
  \\
 F^{\gamma}_{2RL} &=& F^{\gamma}_{2LR} ( h
  \leftrightarrow h', R \leftrightarrow L),
\end{eqnarray}
with $x_{ka} = m^2_{(u^c_a)} / M^2_{S_k}$. 
%and the expressions of the kinematic
%loop functions $F_{i}$($i=1,...,4$) are given in the Appendix B. 
The branching ratio of
$\ell \to \ell' \gamma$ is given by:
\begin{eqnarray}
{\rm Br}(\ell \to \ell' \gamma) &=&\frac{\alpha_{em}}{4
\Gamma(\ell)} \frac{(m^2_\ell - m^2_{\ell'})^3}{ m^3_\ell } \bigg(
|F^{\gamma}_{2LR}|^2 + |F^{\gamma}_{2RL}|^2 \bigg),
\end{eqnarray}
%------------------------------------
In our numerical calculations we analyze the Brs of the decays
under consideration by using the total decay widths of the
decaying leptons $\Gamma(\ell)$.
%===============================================

%===============================================
\section{$\ell^- \to \ell^-_i \ell^-_j \ell^+_j$}
%============================================
In this section, we present the analytical results for the LFV
$\tau$ decay into three leptons with different flavor within
leptoquark model. Next, we  give the analytical results relative
to the branching ratios of $\tau^- \to \ell^-_i \ell^-_j \ell^+_j$
(the analogous results in the muon sector can be obtained by means
of a simple generalization.)
We perform a complete one-loop calculation of the $\tau$ decay
width for all six possible channels, $\tau^- \to \mu^- \mu^-
\mu^+$, $\tau^- \to e^- e^- e^+$, $\tau^- \to \mu^- \mu^+ e^-$,
$\tau^- \to e^- e^+ \mu^-$, $\tau^- \to \mu^- \mu^- e^+$ and
$\tau^- \to \mu^+ e^- e^-$. The  contribution generated by the
$\gamma$-, Z-penguins and box diagrams are presented here
separately. Throughout this section we follow closely the notation
and thr way of presentation of \cite{Hisano:1995cp}.

First, we define the amplitude for $\tau^-(p) \to \ell^-_i
(p_1)\ell^-_j(p_2) \ell^+_j (p_3)$ decays as the sum of the
various contributions,
\begin{eqnarray}
\label{tot} {\mathcal{A}}(\tau^- \to \ell^-_i \ell^-_j \ell^+_j) =
{\mathcal{A}}_{\gamma -penguin} + {\mathcal{A}}_{Z-penguin} +
{\mathcal{A}}_{box}.
\end{eqnarray}
In the following subsections, we present the results for these
contributions in terms of some convenient form factors.
%**************************************************************
%****              Figure 2
%*******************************************************
\begin{figure}
\begin{center}
\vspace{3cm}
%\FIGURE{
\input{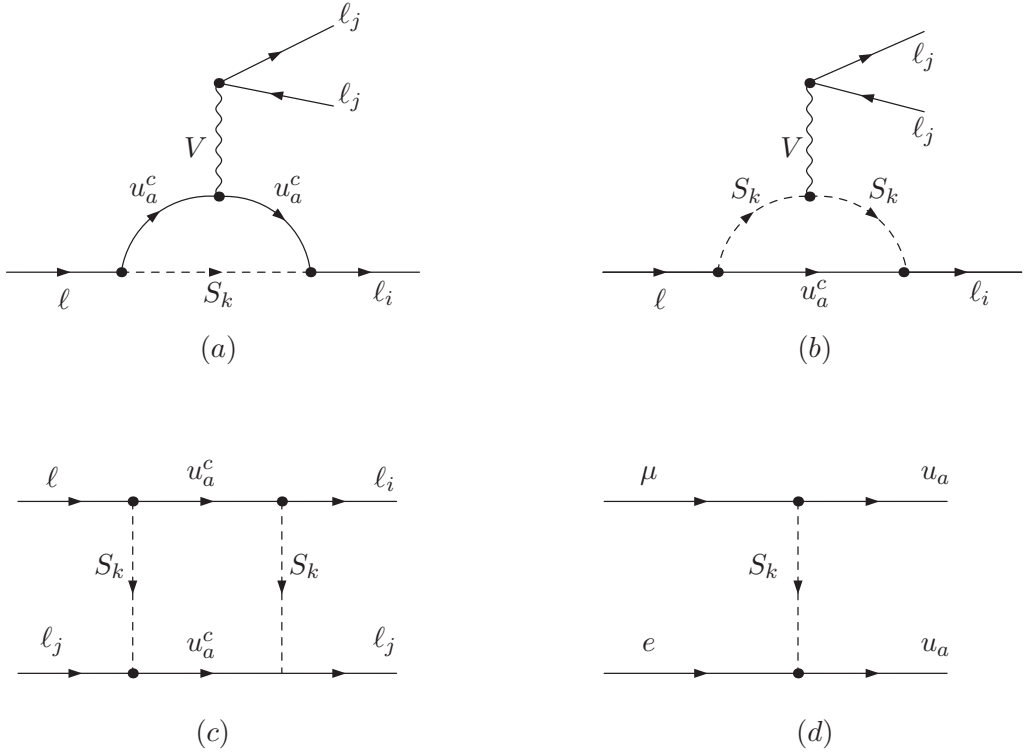}
\vspace{-8.2cm}
%\input{box_LQ.tex}
%\vspace{-14.2cm}
\caption{ Photon (a) and Z-penguin (b) and  box (c) Feynman
diagrams contributing to  $\ell^- \to \ell^-_i \ell^-_j \ell^+_j$,
${S}_k$ are the scalar leptoquark $k=1,2$, $u^c_a$ are type-up quark
with  $a=1,2,3$. The (d) $(\mu-e)$ conversion Feynman diagram. } 
\label{fig:tau3m-diagrams} 
%}
\end{center}
\end{figure}
%===============================================
%====================================================
\subsection{The $\gamma$-penguin contributions}
%========================================================
Diagrams in which  a photon is exchanged are referred as
$\gamma$-penguin diagrams and are shown in
Figs.~\ref{fig:tau3m-diagrams}(a) and \ref{fig:tau3m-diagrams}(b) when 
$V=\gamma$. The
amplitude of $\tau^-(p) \to \ell^-_i(p_1) \ell^-_j(p_2)
\ell^+_j(p_3)$ decays can be written as
%%%%%%%%%%%%%%%%%%%%%%%%%%55
\begin{eqnarray}
\label{ampgez} i{\mathcal{A}}_{\gamma- penguin}&=& \bar{u}(p_1)
\Big[ q^2 \gamma_\mu (T^L_1 P_L + T^R_1 P_R) + i m_{\tau}
\sigma_{\mu\nu} q^\nu (T^L_2 P_L + T^R_2 P_R)\Big]u(p) \\ \no
&\times& \frac{e^2}{q^2} \bar{u}(p_2) \gamma^\mu v(p_3),
\end{eqnarray}
%%%%%%%%%%%%%%%%%%%55
where $q$ is the photon momentum and $e$ is the electric charge.
The photon-penguin amplitude has two contributions, one from
Fig.~\ref{fig:tau3m-diagrams}(a) and the other from
Fig.~\ref{fig:tau3m-diagrams}(b) diagrams respectively as can be
seen from the structure of the form factors,

\begin{eqnarray}
T^{L,R}_{i} =  T^{(a)L,R}_{i} + T^{(b)L,R}_{i}, \qquad {\rm i =
1,2}
\end{eqnarray}
\begin{eqnarray}
 T^{(a)L}_{1} &=&-\frac{N_c Q_{(u^c)}}{16 \pi^2}\sum_{a=1}^{3} \sum_{k=1}^{2}\frac{1}{M^2_{S_k}} h^{\prime}_{a\tau} h^{\prime *}_{ai} \Gamma^{\dagger}_{S_R, k} \Gamma_{k,S_R} F_5 (x_{ka}),\\
%\end{eqnarray}
%
%\begin{eqnarray}
 T^{(a)L}_{2} &=& - \frac{N_c Q_{(u^c)}}{16 \pi^2}\sum_{a=1}^{3} \sum_{k=1}^{2}\frac{1}{M^2_{S_k}}\Bigg[ h_{a\tau} h^*_{ai} \Gamma^{\dagger}_{S_L,k} \Gamma_{k, S_L}  F_1 (x_{ka}) + h^{\prime}_{a\tau} h^{\prime *}_{ai} \Gamma^{\dagger}_{S_R,k} \Gamma_{k, S_R} \frac{m_{i}}{m_{\tau}} F_1 (x_{ka}) \nonumber\\
 &+& h^{\prime}_{a\tau} h^{*}_{ai} \Gamma^{\dagger}_{S_L, k} \Gamma_{k,S_R} \frac{m_{u_a}}{m_\tau}F_3 (x_{ka})\bigg]\\
T^{(a)R}_{i} &=& T^{(a)L}_{i} (h \leftrightarrow h', R
\leftrightarrow L).
\end{eqnarray}
and,
\begin{eqnarray}
 T^{(b)L}_{1} &=&-\frac{N_c Q_S}{16 \pi^2}\sum_{a=1}^{3} \sum_{k=1}^{2}\frac{1}{M^2_{S_k}} h^{\prime}_{a\tau} h^{\prime *}_{ai} \Gamma^{\dagger}_{S_R, k} \Gamma_{k,S_R} F_6 (x_{ka}),\\
%\end{eqnarray}
%
%\begin{eqnarray}
 T^{(b)L}_{2} &=&\frac{N_c Q_S}{16 \pi^2}\sum_{a=1}^{3} \sum_{k=1}^{2}\frac{1}{M^2_{S_k}}\Bigg[ h_{a\tau} h^{*}_{ai} \Gamma^{\dagger}_{S_L,k} \Gamma_{k, S_L}  F_2 (x_{ka}) + h^{\prime}_{a\tau} h^{\prime *}_{ai} \Gamma^{\dagger}_{S_R,k} \Gamma_{k, S_R}  \frac{m_{i}}{m_{\tau}} F_2 (x_{ka}) \nonumber\\ &+& h^{\prime}_{a\tau} h^{*}_{ai} \Gamma^{\dagger}_{S_L,k}\Gamma_{k, S_R}  \frac{m_{u_a}}{m_\tau}F_4 (x_{ka})\bigg]\\
T^{(b)R}_{i} &=& T^{(b)L}_{i} (h \leftrightarrow h', R
\leftrightarrow L).
\end{eqnarray}
where $x_{ka} = m^2_{u_a}/M^2_{S_k}$. Note that we have not
neglected any of the fermion masses. The analytical expressions for the 
loop functions  $F_{i}$ ($i = 1,...,6$) are given in  appendix B.

%====================================================

%====================================================
\subsection{The $Z$-penguin contributions}
%========================================================

In addition to the photon penguin diagrams discussed in the
previous subsection, there are other types of penguin diagrams in
which the $Z$ boson is exchanged as shown in
Figs.~\ref{fig:tau3m-diagrams}(a)-\ref{fig:tau3m-diagrams}(b). The amplitude in this case
can be written as

\begin{eqnarray}
\label{ampge1} i{\mathcal{A}}_{Z- penguin}&=& \frac{i e^2}{m_Z^2
c^2_W s^2_W}\bar{u}(p_1) \gamma_\mu \big(Z^L P_{L} + Z^R P_R\big)
u(p)\\\no  &\times& \bar{u}(p_2) \gamma^\mu \big(g_{L} P_{L} + g_R
P_R\big) v(p_3) ,
\end{eqnarray}
As before, the coefficient $Z^{L(R)}$ can be written as a sum of
 two terms from Feynman diagrams in
Fig.~\ref{fig:tau3m-diagrams}(a) and
Fig.~\ref{fig:tau3m-diagrams}(b):
\begin{eqnarray}
Z^{L,R} = Z^{(a)L,R} + Z^{(b)L,R}
\end{eqnarray}
where,
\begin{eqnarray}
\label{Zcou}
 Z^{(a)L} &=&-\frac{N_c}{16 \pi^2}\sum_{a=1}^{3} \sum_{k=1}^{2}\frac{1}{M^2_{S_k}} h^{\prime}_{a\tau} h^{\prime *}_{ai} \Gamma^{\dagger}_{S_R,k} \Gamma_{k, S_R} \bigg[ 2 C_{R} F_8(x) - m^2_{u_a} C_{L} F_{7}(x_{ka})\bigg],\\
Z^{(a)R} &=& Z^{(a)L}(h^\prime \rightarrow h, R \leftrightarrow L).\\
Z^{(b)L} &=&-\frac{N_c}{16 \pi^2}\sum_{a=1}^{3} \sum_{k=1}^{2}\frac{1}{M^2_{S_k}} h^{\prime}_{a\tau} h^{\prime *}_{ai}\Gamma^{\dagger}_{S_R, k} \Gamma_{k, S_R} \bigg[ 2 Q_S \tan\theta_W  \bigg]F_{8}(x_{ka}),\\
Z^{(b)R} &=& Z^{(b)L}(h^\prime \rightarrow h, R \leftrightarrow
L).
\end{eqnarray}
the coefficients $C_{L(R)}$ and $g_{L(R)}$ denote Z boson coupling
to charged leptoquark S and charged leptons $l_{L(R)}$,
respectively and they are given by
\begin{eqnarray}
g_{L(R)} &=& T_{3L(R)} - Q_{em} \sin^2\theta_W,\\
C_{L(R)} &=& T_{3L(R) (u^c)} - Q_{(u^c)} \sin^2\theta_W,
\end{eqnarray}
where $T_{3L(R)}$ and $Q_{em}$ represent weak isospin and electric
charge of $l_{L(R)}$, respectively. The loop functions $F_i$ (i=7,8) are presnted
in the appendix B.
%====================================================
\subsection{The box contribution}
%========================================================
The amplitude corresponding to the box-type diagram shown in Fig~.\ref{fig:tau3m-diagrams}(c) 
can be expressed as,
\begin{eqnarray}
\label{ampge21} i{\mathcal{A}}_{box}\no &=& B^L_1 [\bar{u}(p_1)
\gamma^\mu P_L u(p)][\bar{u}(p_2) \gamma_\mu P_L v(p_3)] + B^R_1
[\bar{u}(p_1) \gamma^\mu P_R u(p)][\bar{u}(p_2) \gamma_\mu P_R
v(p_3)] \\ \no &+& B^L_2 [\bar{u}(p_1) \gamma^\mu P_L
u(p)][\bar{u}(p_2) \gamma_\mu P_R v(p_3)] + B^R_2 [\bar{u}(p_1)
\gamma^\mu P_R u(p)][\bar{u}(p_2) \gamma_\mu P_L v(p_3)]\\\no&+&
B^L_3 [\bar{u}(p_1) P_L u(p)][\bar{u}(p_2) P_L u(p)] + B^R_3
[\bar{u}(p_1) P_R u(p)][\bar{u}(p_2) P_R v(p_3)]\\\no &+& B^L_4
[\bar{u}(p_1) \sigma^{\mu \nu}P_L u(p)][\bar{u}(p_2) \sigma_{\mu
\nu} P_L v(p_3)] \\ &+& B^R_4 [\bar{u}(p_1) \sigma^{\mu \nu} P_R
u(p)][\bar{u}(p_2) \sigma_{\mu \nu} P_R v(p_3)].
\end{eqnarray}
where
\begin{eqnarray}
B^{L,R}_i = B^{(c)L,R}_i  \qquad i = 1,...,4
\end{eqnarray}
with,
\begin{eqnarray}
B^{(c)L}_1 &=& \frac{N_c}{32\pi^2} \sum^3_{a,a'=1}\sum^{2}_{k,k'=1} \widetilde{D}_0 (m^2_{u_a},m^2_{u_{a'}}, m^2_{S_k}, m^2_{S_{k^\prime}}) h^{\prime}_{a\tau} h^{\prime}_{a'j}h^{\prime *}_{a i} h^{\prime *}_{a' j}| \Gamma^\dagger_{S_R,k} \Gamma_{k^\prime,S_R}|^2 ,\\
B^{(c)L}_2 &=& \frac{N_c}{64\pi^2} \sum^3_{a,a'=1}\sum^{2}_{k,k'=1} h^{\prime}_{a \tau}h_{a'j}  \Gamma_{k,S_R} \Gamma_{k^\prime,S_L} \bigg[h^{*}_{a'j}  h^{\prime *}_{ai}\Gamma^\dagger_{S_R,k'} \Gamma^\dagger_{S_L,k} \widetilde{D}_0 (m^2_{u_a},m^2_{u_{a'}}, m^2_{S_k}, m^2_{S_{k^\prime}})   \nonumber\\ && - m_{u_a} m_{u_{a'}} h^{\prime *}_{a^\prime j}  h^{*}_{ai}\Gamma^\dagger_{S_R,k} \Gamma^\dagger_{S_L,k^\prime} D_0 (m^2_{u_a},m^2_{u_{a'}}, m^2_{S_k}, m^2_{S_{k^\prime}})\bigg],\\
B^{(c)L}_3 &=& \frac{N_c}{16\pi^2}
\sum^3_{a,a'=1}\sum^{2}_{k,k'=1}  m_{u_a} m_{u_a'}
h'_{a'j}h'_{a\tau} h^{*}_{ai}  h^{*}_{a'j} \Gamma^\dagger_{S_L,k'}
\Gamma_{k,S_R}\Gamma^\dagger_{S_L,k} \Gamma_{k',S_R}\nonumber\\ &&
\,\,\,\,\,\,\,\,\,\,\,~~~~~~~~~~~~~~~~~~~~~~~~~~~~~~~~~~~~~~~~~~~~\times
D_0(m^2_{u_a},m^2_{u_{a'}}, m^2_{S_k}, m^2_{S_{k'}})
\\
B^{(c)L}_4 &=& 0,\\
B^{(c)R} &=& B^{(c)L}(h^\prime \leftrightarrow h, R
\leftrightarrow L).
\end{eqnarray}
Again the loop functions $D_0$ and $\widetilde{D}_0$ are given in the appendix B.\\
By collecting all the formulas, the Branching ratios of $\tau^- \to \ell^-_i \ell^-_j \ell^+_j$ can
be written in terms of the different form factors as
\begin{eqnarray}
\rm{Br}(\tau^- \to \ell^-_i \ell^-_j \ell^+_j)& = &\no
\frac{\alpha^2 m^5_{\tau}}{32 \pi \Gamma_\tau}  \Bigg[ |T^L_1|^2 +
|T^R_1|^2 + \frac{2}{3}\bigg(|T^L_2|^2 + |T^R_2|^2\bigg)
\bigg(8\log\bigg(\frac{m_{\tau}}{2m_i}\bigg) {- 11}\bigg)\\\no
&-&2 (T^L_1 T^{R*}_2 + T^L_2 T^{R*}_1 + {\rm h.c}) + {\frac{1}{3
m^4_Z s^4_W c^4_W}}\bigg({2}\big(|Z^L g_L|^2 + |Z^R g_R|^2\big)
\\\no &+& |Z^L g_R|^2 + |Z^R g_L|^2 \bigg) + \frac{1}{6}
\big(|B^L_1|^2 + |B^R_1|^2) + \frac{1}{3} \big(|B^L_2|^2 +
|B^R_2|^2)
\\\no &+& \frac{1}{24} \big(|B^L_3|^2 + |B^R_3|^2) + \frac{1}{3}\big( T^L_1 B^{L*}_1 + T^L_1 B^{L*}_2 + T^R_1 B^{R*}_1 + T^R_1 B^{R*}_2 +{\rm h.c}\big) \\\no &-& \frac{2}{3}\big( T^R_2 B^{L*}_1 + T^L_2 B^{R*}_1 + T^L_2 B^{R*}_2 + T^R_2 B^{L*}_2 +{\rm h.c}\big)\\\no &+& \frac{1}{3}\big( B^L_1 Z^{*}_L g_L + B^R_1 Z^{*}_R g_R + B^L_2 Z^{*}_L g_R + B^R_2 Z^{*}_R g_L + {\rm h.c}\big)\\\no &+& \frac{1}{3}\big[ 2( T^L_1 Z^{*}_L g_L + T^R_1 Z^{*}_R g_R)  + T^L_1 Z^{*}_L g_R + T^R_1 Z^{*}_R g_L + {\rm h.c}\big]\\ &+& \frac{1}{3}\big[ -4( T^R_2 Z^{*}_L g_L + T^L_2 Z^{*}_R g_R) {-2 (T^L_2 Z^{*}_R g_L + T^R_2 Z^{*}_L g_R + {\rm h.c})}\big] \Bigg]
\end{eqnarray}
where $\Gamma_\tau$ is the total decay width of $\tau$. All the form factors are real.

%=======================================================

%=======================================================
\section{$\mu -e $ conversion }
%==================================
%
%\begin{figure}[t]
%\begin{center}
%\includegraphics[width=4cm]{mue.eps}
%\vskip -0.3cm \caption{Tree-level diagram contributing to the
%$\mu$-e conversion}
%\end{center} \label{mue}
%\end{figure}
%====================================
 $\mu-e$ conversion in the
muonic atoms is one of the interesting charged LFV process that
can occur in many candidates of physics beyond the SM. Accurate
calculation of the $\mu- e$ conversion rate is essential to
compare the sensitivity to the LFV interactions in different
nuclei~\cite{Kitano:2002mt}.
 In this section, we discuss the constraints that can be imposed  on
 the scalar leptoquark couplings using  $\mu-e$ conversion
 rate.  The dominant contribution to the $\mu- e$ conversion rate
 is obtained through considering the tree diagram shown in
 Fig.~\ref{fig:tau3m-diagrams}(d) which leads to the  effective Lagrangian
\bea
 \mathcal L_{eff}^{(u_a)} & = & \sum_{a=1}^{3} \sum_{k=1}^{2}
-\frac{1}{M^2_{S_k}}\Bigg[ \frac{1}{2} h_{a2} h^{*}_{a1}
\Gamma^{\dagger}_{S_L,k} \Gamma_{k,S_L} (\bar{e}\mathcal
\gamma^{\mu}P_L \mu )(\bar{u}_a\mathcal \gamma _{\mu}P_L u_a )
\nonumber\\&+&\frac{1}{8} h_{a2} h^{'*}_{a1}
\Gamma^{\dagger}_{S_R,k} \Gamma_{k,S_L} (\bar{e}\sigma^{\mu\nu}P_R
\mu )(\bar{u}_a \sigma_{\mu\nu}P_R u_a) \nonumber\\&-&\frac{1}{2}
h_{a2} h^{'*}_{a1} \Gamma^{\dagger}_{S_R,k} \Gamma_{k,S_L}
(\bar{e}P_R \mu )(\bar{u}_a P_R u_a )+ (h \leftrightarrow h', R
\leftrightarrow L)\Bigg], \eea
where we have used  Fierz transformation for chiral fermions.
$P_{R,L}=(1\pm \gamma^5) / 2$, $u_{a}$ are light and heavy type-up
quarks and  $\sigma$ matrix is defined by
$\sigma^{\mu\nu}=\frac{i}{2}[\gamma^{\mu},\gamma^{\nu}]$. The
 operators involving $\bar{u}_a \gamma_\mu \gamma_5 u_a$,
$\bar{u}_a  \gamma_5 u_a$,  or  $\bar{u}_a \sigma_{\mu \nu}  u_a$
do not contribute to the coherent conversion processes and thus we
can drop them and write
\begin{eqnarray}
\mathcal L_{eff}^{(u_a)} & = & \sum_{a=1}^{3} \ \Bigg[  \left(
C^{(u_a)}_{VR}  \; \bar{e}\mathcal \gamma^{\mu}P_R \mu  +
C^{(u_a)}_{VL}  \; \bar{e}\mathcal \gamma^{\mu} P_L \mu \right)
\bar{u}_a\gamma_{\mu}u_a
\nonumber\\
&+&  \left( C^{(u_a)}_{SR}  \; \bar{e} P_L \mu + C^{(u_a)}_{SL}
 \; \bar{e} P_R \mu \right) \bar{u}_a u_a  ~ \Bigg] ~.
\label{eq:weakscaleL}
\end{eqnarray}
where we have defined \bea C^{(u_a)}_{VR}&=&- h_{a2}
h^*_{a1}\sum_{k}\frac{1}{2
M^2_{S_K}} \Gamma^{\dagger}_{S_L,k} \Gamma_{k,S_L}\nonumber\\
C^{(u_a)}_{SR}&=& \frac{1}{2}h_{a2} h^{'*}_{a1} \sum_{k}\frac{1}{
M^2_{S_K}} \Gamma^{\dagger}_{S_R,k}
\Gamma_{k,S_L}\label{coeff}\eea
$C^{(u_a)}_{VL}$ and $C^{(u_a)}_{SL}$ can be obtained by the
the exchange $h \leftrightarrow h', R \leftrightarrow L$ in
Eq.(\ref{coeff}). 
The next step for the calculation of $\mu-e$ conversion  is
to match the Lagrangian in Eq.(\ref{eq:weakscaleL}) to the
Lagrangian at the nucleon level.   
Hence  we integrate out the
heavy quarks~\cite{Cirigliano:2009bz} and so the effective
Lagrangian in Eq.(\ref{eq:weakscaleL}) becomes
\begin{eqnarray}\label{lagrangian}
\mathcal L_{\rm eff}^{(u)}&=&
 \left( C^{(u)}_{VR}  \; \bar{e}\mathcal
\gamma^{\mu}P_R \mu  +
 C^{(u)}_{VL}  \; \bar{e}\mathcal \gamma^{\mu} P_L \mu \right)
 \bar{u}\gamma_{\mu}u
 %\right.
 \nonumber\\
 &+&
% \left.
 \left( C^{(u)}_{SR}  \; \bar{e} P_L \mu
 + C^{(u)}_{SL}   \; \bar{e} P_R \mu \right)
 \bar{u}u.
\end{eqnarray}

Then, the effective Lagrangian (\ref{lagrangian}) is matched
to the nucleon level Lagrangian~\cite{Kosmas:2001mv} through 
the following replacements of the operators~\cite{Kitano:2002mt,Cirigliano:2009bz}:
 \begin{eqnarray}
  \bar{u}u  &\rightarrow& G_{S}^{(u,N)}
{\bar{\psi}}_N {\psi}_N
\nonumber\\
\bar{u}\gamma_{\mu} u  &\rightarrow& f_{VN}^{(u)} \;
{\bar{\psi}}_N \gamma_{\mu} {\psi}_N \;, \label{eq:replace}
\end{eqnarray}
where $N$ represents each nucleon ($N=p,n$), $\psi_N$ are the
nucleon fields, and $G,f$ are given by~\cite{Kitano:2002mt,Cirigliano:2009bz}
\begin{equation}
 f_{Vp}^{(u)}=2,\,\,\,\,\,\,\, \,\,\,\, f_{Vn}^{(u)}=1,\,\,\,\,\,
 \,\,G_{S}^{(u,p)}= 5.1,\,\,\,\,  \,\,G_{S}^{(u,n)}= 4.3
\end{equation}
Finally, the Lagrangian at nucleon level can be written as
\begin{eqnarray}
\mathcal L_{eff}^{(N)}&=& \sum_{N=p,n} \Bigg[ \left(
\tilde{C}^{(N)}_{VR}  \; \bar{e} \gamma^{\mu} P_R \mu  +
\tilde{C}^{(N)}_{VL} \; \bar{e} \gamma^{\mu} P_L \mu \right)  \
\bar{\psi}_N \gamma_{\mu} \psi_N
%\right.
\nonumber\\
&+&
%\left.
    \, \left( \tilde{C}^{(N)}_{SR}  \; \bar{e} P_L
\mu +  \tilde{C}^{(N)}_{SL}  \; \bar{e} P_R \mu \right) \;
\bar{\psi}_N \psi_N +  h.c.\Bigg] \; .
\end{eqnarray}
 where we have introduced the following redefinitions for the vector quantities:
\begin{eqnarray}
\tilde{C}^{(p)}_{VR} &=& C^{(u)}_{VR} \; f^{(u)}_{Vp} \\
\tilde{C}^{(n)}_{VR} &=&  C^{(u)}_{VR} \; f^{(u)}_{Vn} \\
\tilde{C}^{(p)}_{VL} &=&  C^{(u)}_{VL} \; f^{(u)}_{Vp} \\
\tilde{C}^{(n)}_{VL} &=&   C^{(u)}_{VL} \; f^{(u)}_{Vn} ~,
\end{eqnarray}
while the scalar ones read:
\begin{eqnarray}
\tilde{C}^{(p)}_{SR} &=&   C^{(u)}_{SR} \; G^{(u,p)}_{S}
 \\
\tilde{C}^{(n)}_{SR} &=&  C^{(u)}_{SR} \; G^{(u,n)}_{S}
 \\
\tilde{C}^{(p)}_{SL} &=&  C^{(u)}_{SL} \;
G^{(u,p)}_{S}\\
\tilde{C}^{(n)}_{SL} &=&   C^{(u)}_{SL} \; G^{(u,n)}_{S} \ .
\end{eqnarray}

In order to calculate the $\mu - e$ conversion amplitude we need
to calculate the  matrix elements of $\bar{\psi}_N \psi_N $ and
$\bar{\psi}_N \gamma_{\mu}\psi_N $ of the transition between the
initial and the final states of nucleus~\cite{Kitano:2002mt,Cirigliano:2009bz}:
\begin{eqnarray}
    \langle A,Z|\bar{\psi}_p\psi_p|A,Z\rangle&=&Z\rho^{(p)} \nonumber\\
    \langle A,Z|\bar{\psi}_n\psi_n|A,Z\rangle&=&(A-Z)\rho^{(n)} \nonumber\\
    \langle A,Z|\bar{\psi}_p\gamma^0\psi_p|A,Z\rangle &=&Z\rho^{(p)} \nonumber\\
    \langle A,Z|\bar{\psi}_n\gamma^0\psi_n|A,Z\rangle &=&(A-Z)\rho^{(n)} \nonumber\\
    \langle A,Z|\bar{\psi}_N\gamma^i\psi_N|A, Z \rangle &=&0 \; .
\end{eqnarray}
where $|A,Z\rangle$ represents the nuclear ground state, with $A$
and $Z$ are the mass and atomic number of the isotope
respectively, while $\rho^{(p)}$ and $\rho^{(n)}$ are the proton
and neutron densities respectively.  
Finally, the $\mu-e $
conversion rate is given by~\cite{Cirigliano:2009bz}:
\begin{eqnarray}
\Gamma_{conv} &=& \frac{m_{\mu}^{5}}{4} \left|  4 \left(
\tilde{C}^{(p)}_{SR}  S^{(p)} +  \tilde{C}^{(n)}_{SR} \; S^{(n)}
\right) + 4\tilde{C}^{(p)}_{VR} \; V^{(p)} + 4\tilde{C}^{(n)}_{VR}
\;V^{(n)} \right|^2
\nonumber \\
&+&  \frac{m_{\mu}^{5}}{4} \left|  4 \left( \tilde{C}^{(p)}_{SL}
S^{(p)} +  \tilde{C}^{(n)}_{SL} \; S^{(n)} \right) +
4\tilde{C}^{(p)}_{VL} \;V^{(p)} + 4\tilde{C}^{(n)}_{VL} \;
V^{(n)} \right|^2
\end{eqnarray}
where $ V^{(N)}, S^{(N)}$ are  dimensionless integrals
representing the overlap of electron and muon wave functions
weighted by appropriate combinations of protons and neutron
densities~\cite{Kitano:2002mt}. For phenomenological applications,
it is useful to normalize the conversion rate to the muon capture
rate through the quantity:
\begin{equation}
B_{\mu- e} (Z)  \equiv   \frac{ \Gamma_{conv}(Z,A)
}{\Gamma_{capt} (Z,A) }~.
\end{equation}
 The current bounds on $B_{\mu- e}  $ for   Titanium  atom  and Gold atom obtained by
 SINDRUM collaboration  are respectively $B_{\mu- e} (Ti)< 4.3\times 10^{-12}$~\cite{Dohmen:1993mp},  $B_{\mu- e} (Au)< 7\times 10^{-13}$
~\cite{Bertl:2006up} both at 90\%CL.
The numerical values  of $ V^{(N)}, S^{(N)}$ 
and $\Gamma_{capt}$ for Titanium and Gold atoms are listed
In  Table~\ref{TabTiAuData}.

%%%%%%%%%%%%%%%%%%%%%%%%%%%%%%%%%%%%%%%%%%%%%%%%%%%%%%%%%%%%%%%555
\begin{table}
  \begin{center}
\begin{tabular}{|c|c|c|c|c|c|}
  \hline
   Nucleus  & $S^{(p)} [m_\mu^{5/2}]$ & $S^{(n)} [m_\mu^{5/2}]$
   & $V^{(p)} [m_\mu^{5/2}]$& $V^{(n)} [m_\mu^{5/2}]$& $\Gamma_{capture}[10^6 s^{-1}]$ \\
  \hline
   $\text{Ti}^{48}_{22}$ &  0.0368 & 0.0435 &  0.0396 &   0.0468& 2.59 \\
    $\text{Au}^{197}_{79}$& 0.0614 & 0.0918 &  0.0974   &   0.146   & 13.07  \\
      \hline
\end{tabular}
 \end{center}
  \caption{Data taken from Tables I and VIII of \cite{Kitano:2002mt}.}
  \label{TabTiAuData}
\end{table}

%============================================
\section{Numerical results and discussion}
%============================================

Let us now proceed to analyse and discuss our numerical results. 
The quark masses are evaluated  at the energy  scale  $\mu =
300$~GeV~\cite{Gray:1990}, which is the typical leptoquark mass
scale used in this work,
\begin{eqnarray}
m_t = 161.4 \,{\rm GeV}, \quad m_c = 0.55\, {\rm GeV},\quad
 m_u  = 11.4 \times 10^{-3}\,{\rm GeV},
\end{eqnarray}
while we use the following values for ~\cite{pdg}
\begin{eqnarray}
\alpha_{em} = 1/137.0359, \quad M_W = 80.45 \,{\rm GeV} , \quad
M_Z = 91.1875 \,{\rm GeV}.
\end{eqnarray}
%
%motivated by maximal $\nu_{\mu} - \nu_{\tau}$ mixing \cite{CHH1999},
We assume as in Ref.~\cite{Davidson:1993qk}, that all the couplings $h$ and $h'$ are real and equal to each other~\cite{CHH1999},
\begin{eqnarray}
h = h^\prime = h^*.
\end{eqnarray}
We use leptoquark mass splitting $\Delta= 500$ GeV in our
analysis, where $\Delta$ is defined as $\sqrt{M^2_{S_2} -
M^2_{S_1}}$. Consequently, the remaining parameters in the
leptoquark model are the mass of the light scalar leptoquark
$M_{S_1}$, the mixing angle $\theta_{LQ}$, and the couplings
$h_{a\ell}$ (a =u, c and t).  
Also, we assume that the scalar leptoquark may
explain the discrepancy $\Delta a_\mu$ between the experimentally
measured muon $(g-2)_\mu$ and its SM prediction Eq.~(\ref{gm2}) 
and hence this condition restricts the possible range of the
parameters. We have performed a scan over all the input
parameters, $M_{S_1} \le 1500$ GeV, $-1 \le \sin\theta_{LQ} \le 1$
and $\alpha_{em} \le |h_{q\mu}|^2 \le 1 $. Then, after imposing
all the existing constraints arising from $\pi$ leptonic
decays and direct search, we  select all  sets of the input
parameters producing the same values for $(g-2)_\mu$ at 1$\sigma$
range of data.

%=========================================================
%=========================================================
\begin{figure} % fig 1
%\FIGURE{
\begin{picture}(320,260)
\put(-60,0){\mbox{\psfig{file=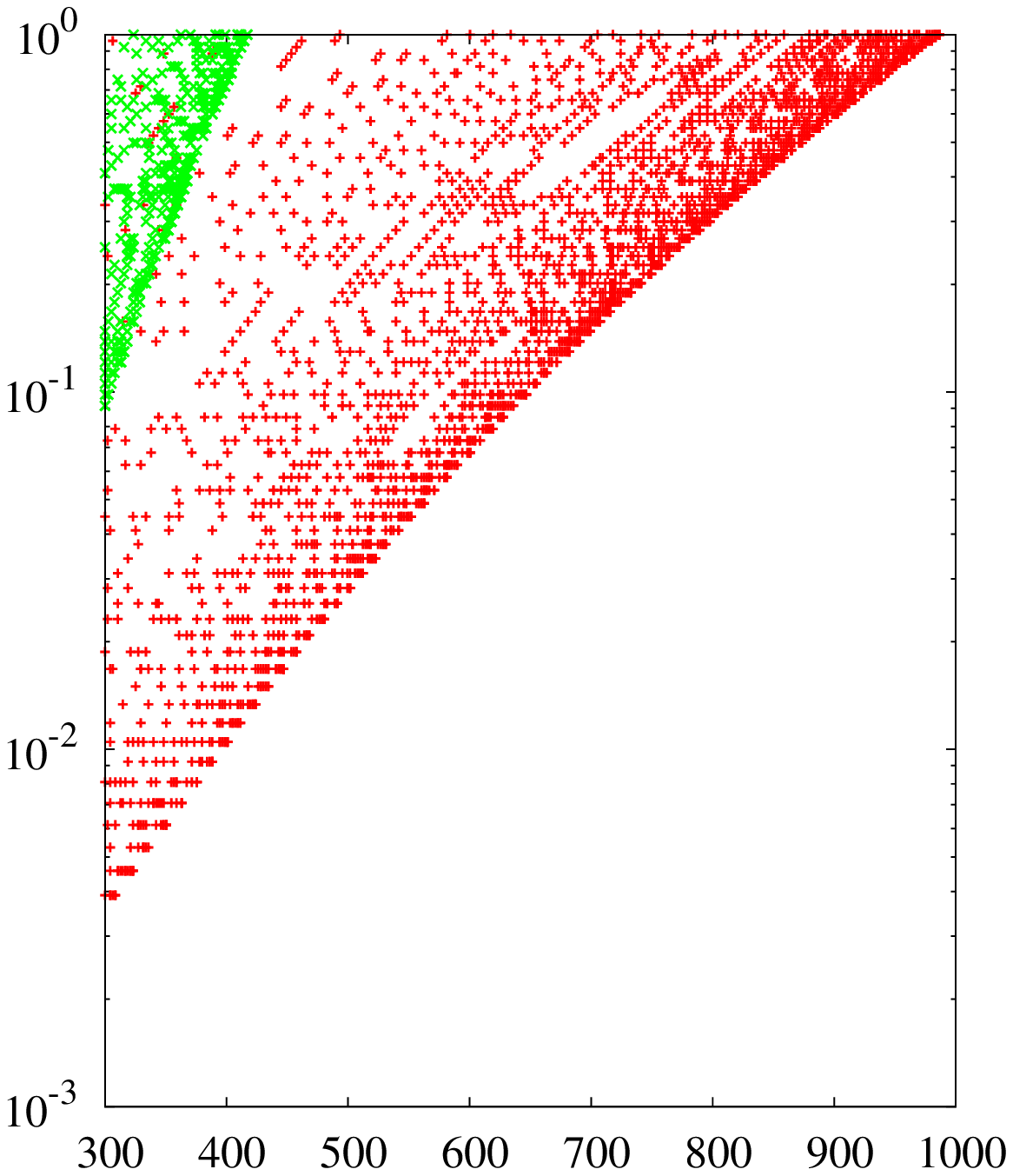,height=3.4in,width=3.2in}}}
\put(33,-10){\makebox(0,0)[bl]{\large{$M_{S_1}$ (GeV)}}}
\put(-60,115){\makebox(0,0)[bl]{\rotatebox{90}{\large{$|h_{a\mu}|^2$
}}}}
\put(-27,170){\makebox(0,0)[bl]{\rotatebox{70}{{Charm-quark
}}}} \put(30,120){\makebox(0,0)[bl]{\rotatebox{45}{{Top-quark}}}}
\put(160,0){\mbox{\psfig{file=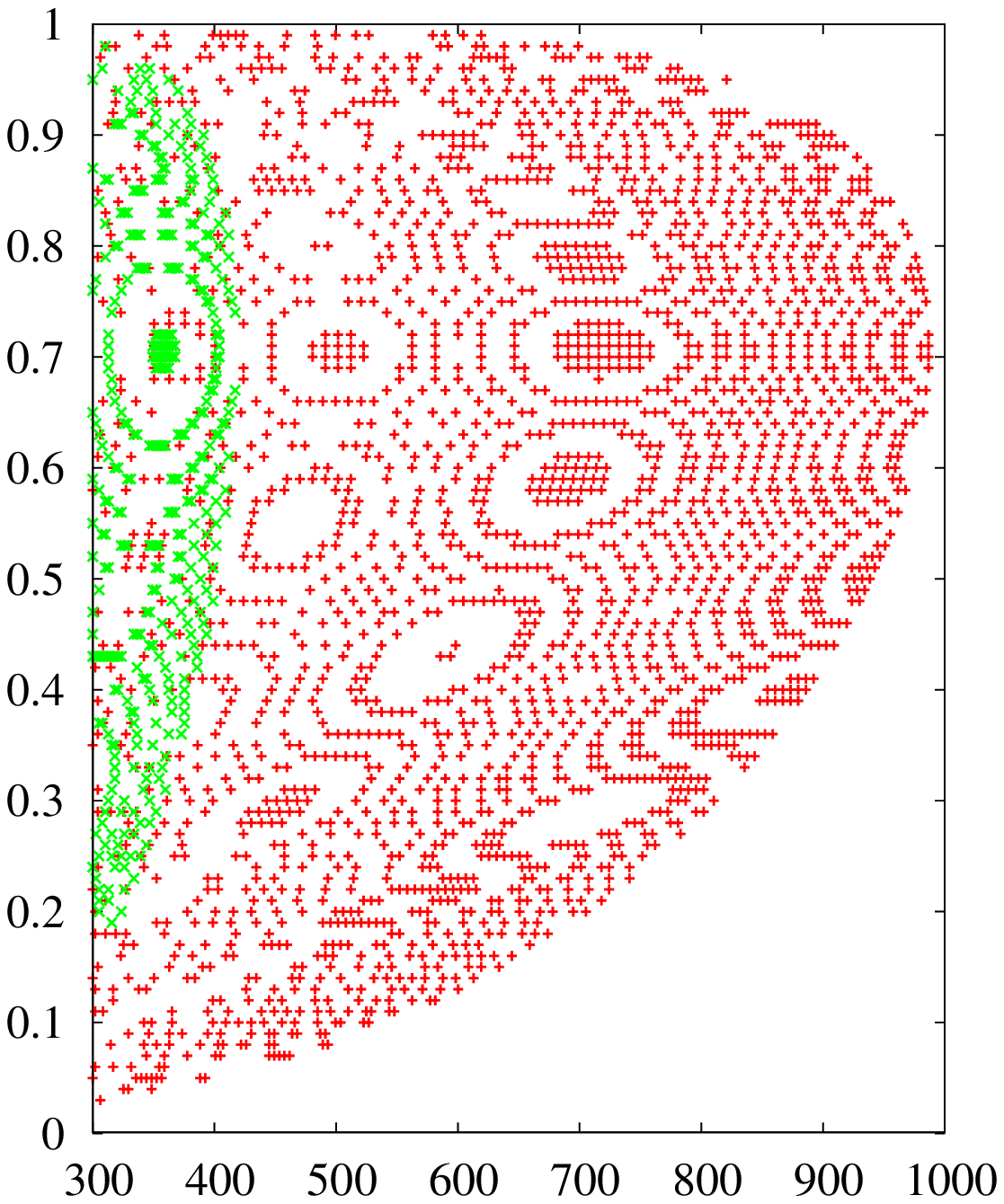,height=3.4in,width=3.2in}}}
\put(165,115){\makebox(0,0)[bl]{\rotatebox{90}{\large{$\sin2\theta_{LQ}$
}}}} \put(253,-10){\makebox(0,0)[bl]{\large{$M_{S_1}$(GeV)}}}
\put(226,30){\makebox(0,0)[bl]{{Top-quark}}}
\put(210,150){\makebox(0,0)[bl]{\rotatebox{90}{{Charm-quark
}}}}
\end{picture}
\vspace{0.6cm} \caption{ The allowed regions on the
($M_{S_1}-|h_{a\mu}|^2$) plane (left) and on the
($M_{S_1}-\sin2\theta_{LQ}$) plane (right) for top-quark (red color) and 
charm-quark (green color) contributions, 
taking into account $a^{LQ}_\mu = \Delta a_{\mu}$
at 1$\sigma$.} \label{fig-gm2}
%}
\end{figure}
%=========================================================
In Fig.~\ref{fig-gm2}, we show the allowed regions for different
type-up quarks  contributions which are  compatible with $a^{LQ}_\mu =
\Delta a_{\mu}$ at 1$\sigma$ range of data where the red color (green color)
region corresponds to top-quark (charm-quark) contribution respectively.
As can be seen from the left panel of Fig.~\ref{fig-gm2}, 
the dominant contribution is
around $\sin2\theta_{LQ} \sim 0.7$ for both  top and charm quarks.
In addition, we find that, sizeable scalar leptoquark effects to
the $(g-2)_\mu$ at 1$\sigma$ are obtained for the values of $M_{S_1}$
which satisfy $M_{S_1} \approx 1$ TeV for top quark contribution and
$M_{S_1} \approx 400$ GeV for charm quark contribution. We note also
that it is not possible to use the up quark loop contributions
alone for  LQ, since the couplings $|h_{u\mu}|$ are strongly
constrained by the $\pi$ leptonic decays. It has been found that the
LHC has the potential to discover light scalar LQ with a mass up to
1.2 TeV and where the Yukawa coupling are equal to the electromagnetic
coupling \cite{Belyaev:2005ew}.

%----------------------------------------------
In order to find the constraints on the combination of LQ
couplings we  require that each individual LQ coupling
contribution to the branching ratio does not exceed the
experimental current limits on the
Br($\ell\to\ell^{\prime}\gamma$) ~(\ref{lllm}) and $\mu-e$
conversion in nuclei. The latter process is used to set the
strongest constraints on the product $h_{u\mu}h_{ue}$ which
involve the first generation, since, in this case, the process is
induced at tree-level. On the other hand, the
$\ell\to\ell^{\prime}\gamma$ decays which are induced at one-loop
by the photon-penguins, Z-penguins and box diagrams [see
Figs.~\ref{fig:tau3m-diagrams}(a)-\ref{fig:tau3m-diagrams}(c)],
allow us to constrain the complementary combinations of the
couplings involving the second and third generation of quarks,
namely $h_{a\ell}h_{a\ell^\prime}$, where a = c, t. The $\mu-e$
conversion process can be also used to set constraints on the
second and third quark generations.
 However, we stress that the bounds from the $\mu-e$ conversion
 are suffering from  being model-dependent due to the non perturbative
calculations of the nuclear form factors, while the bounds which are obtained
from the $\ell\to\ell^{\prime}\gamma$ decays are not. Taking
into account $(g-2)_\mu$ constraint, and experimental upper limits on
the $(\mu-e)_{Ti, Au}$ conversion
rates~\cite{Dohmen:1993mp,Bertl:2006up} at 90\%CL., we obtain the
following upper bounds
\begin{equation}
h_{u\mu}h_{ue} \leq 4.38\times 10^{-6} \label{mue11} \end{equation}
for Titanium atom while for Gold  atom the bound reads
\begin{equation}
h_{u\mu}h_{ue} \leq 6.25\times 10^{-6} \label{mue22}
\end{equation}
Clearly, the bounds obtained for both Titanium and Gold atoms are
of the same order and severely constraint the product of
the leptoquark couplings $h_{u\mu}h_{u\mu}$. Our results are
consistent with the effective Hamiltonian and approximation used
in Ref.~\cite{Davidson:1993qk}. The products of the couplings
$h_{a\ell}h_{a\ell^\prime}$ where a=c,t and $\ell,\ell^\prime =
\tau, \mu, e$, appear  only at the one-loop level contribution to
the $\tau\to (\mu,e) \gamma$ and $\mu\to e\gamma$ decays.
Therefore, they could be larger if they are compared with
$h_{u\mu}h_{ue}$ ones without violating the experimental upper
limits on the branching ratios. The bounds obtained on these
combinations of couplings in the muon sector are given by
\begin{equation}
h_{c\mu}h_{ce} \leq 1.22\times 10^{-3}, \quad  
h_{t\mu}h_{te} \leq 5.73\times 10^{-3}
\end{equation}
while for the tau sector they become
\begin{eqnarray}
h_{c\tau}h_{c\mu} &\leq& 4.78\times 10^{-3}, \quad  
h_{t\tau}h_{t\mu} \leq 8.13\times 10^{-3},\\
h_{c\tau}h_{ce} &\leq& 7.8\times 10^{-1}, \quad  h_{t\tau}h_{te}
\leq 8.0\times 10^{-1},
\end{eqnarray}
 In the following studies of the $\tau$ and $\mu$ LFV processes, 
we will use the parameter
space which is discussed above.  
We start by investigating the LFV $\tau$
and $\mu$ decay processes generated by the same LQ-scalar
interactions as those of $(g-2)_\mu$. At one-loop level, the
LQ-scalar gives contributions to the $\tau^- \to \mu^-
\mu^-\mu^+$, $\tau^- \to e^- e^- e^+$,
%$\tau^- \to e^- e^+ \mu^-$ and $\tau^- \to \mu^- \mu^+ e^-$,
by means of the so-called $\gamma$- and Z- penguins and box diagrams.
%however,
%, $\tau^- \to \mu^- \mu^- e^+$ and
%$\tau^- \to \mu^+ e^- e^-$.
%$\tau^- \to \ell^-_i \ell^-_j \ell^+_j$ decay is
%proceeds through three kinds of Feynman diagrams:
%(i) photon penguin, (ii) Z penguin and (iii) box graph.
%The flavor violation is induced by $h_{q\ell} h_{q\ell^{\prime}}$
%couplings via loops with quarks and scalar LQ as internal lines.
%First we consider  $\tau \to 3 \mu$ and $\tau \to \mu \gamma$.
%Since the constraints obtained on the couplings $h_{q\mu}$ from
%the consistency of $\Delta a_\mu$ in the previous section tolerate
%large LFV couplings. Therefore, we will find  regions of the
%couplings $h_{q\tau}h_{q}\mu$ which are consistent with both
%$\Delta a_\mu$ and LFV tau decays.

%=========================================================
\begin{figure}[H] % fig 2
%\FIGURE{
\begin{picture}(320,260)
\put(10,0){\mbox{\psfig{file=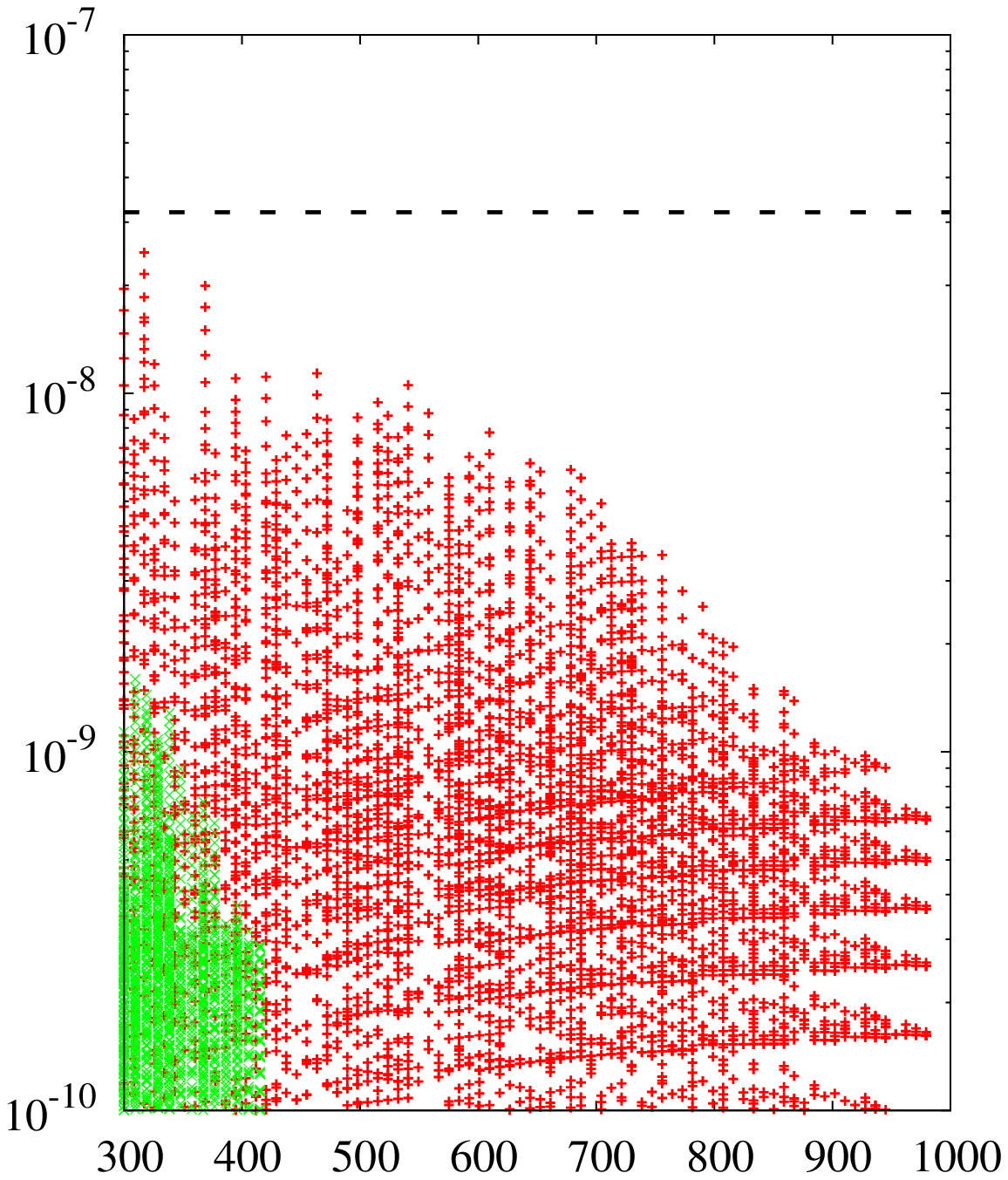,height=3.4in,width=3.2in}}}
\put(173,210){\makebox(0,0)[bl]{\large{ \bf(a)}}}
\put(103,-10){\makebox(0,0)[bl]{\large{$M_{S_1}$ (GeV)}}}
\put(65,203){\makebox(0,0)[bl]{\large{Current limit}}}
\put(10,88){\makebox(0,0)[bl]{\rotatebox{90}{\large{Br$(\tau\to
\mu\mu\bar{\mu})$ }}}}
\put(70,160){\makebox(0,0)[bl]{{Top-quark}}}
\put(55,20){\makebox(0,0)[bl]{\rotatebox{90}{{Charm-quark}}}}
\put(240,0){\mbox{\psfig{file=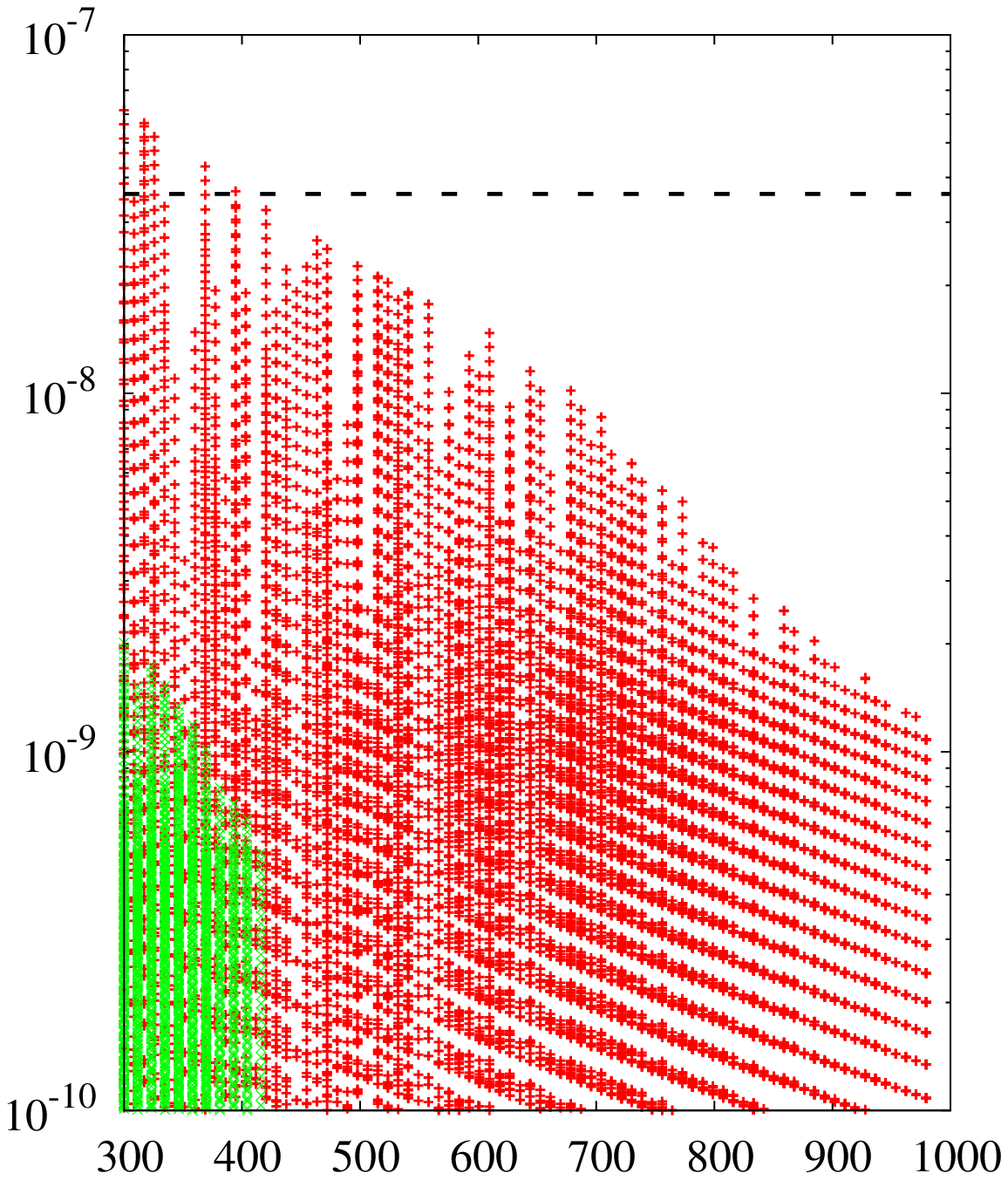,height=3.4in,width=3.2in}}}
\put(404,210){\makebox(0,0)[bl]{\large{ \bf(b)}}}
\put(333,-10){\makebox(0,0)[bl]{\large{$M_{S_1}$(GeV)}}}
\put(240,100){\makebox(0,0)[bl]{\rotatebox{90}{\large{Br$(\tau\to
ee\bar{e})$ }}}}
\put(300,206){\makebox(0,0)[bl]{\large{Current limit}}}
\put(326,170){\makebox(0,0)[bl]{{Top-quark}}}
\put(285,20){\makebox(0,0)[bl]{\rotatebox{90}{{Charm-quark}}}}
\put(120,-290){\mbox{\psfig{file=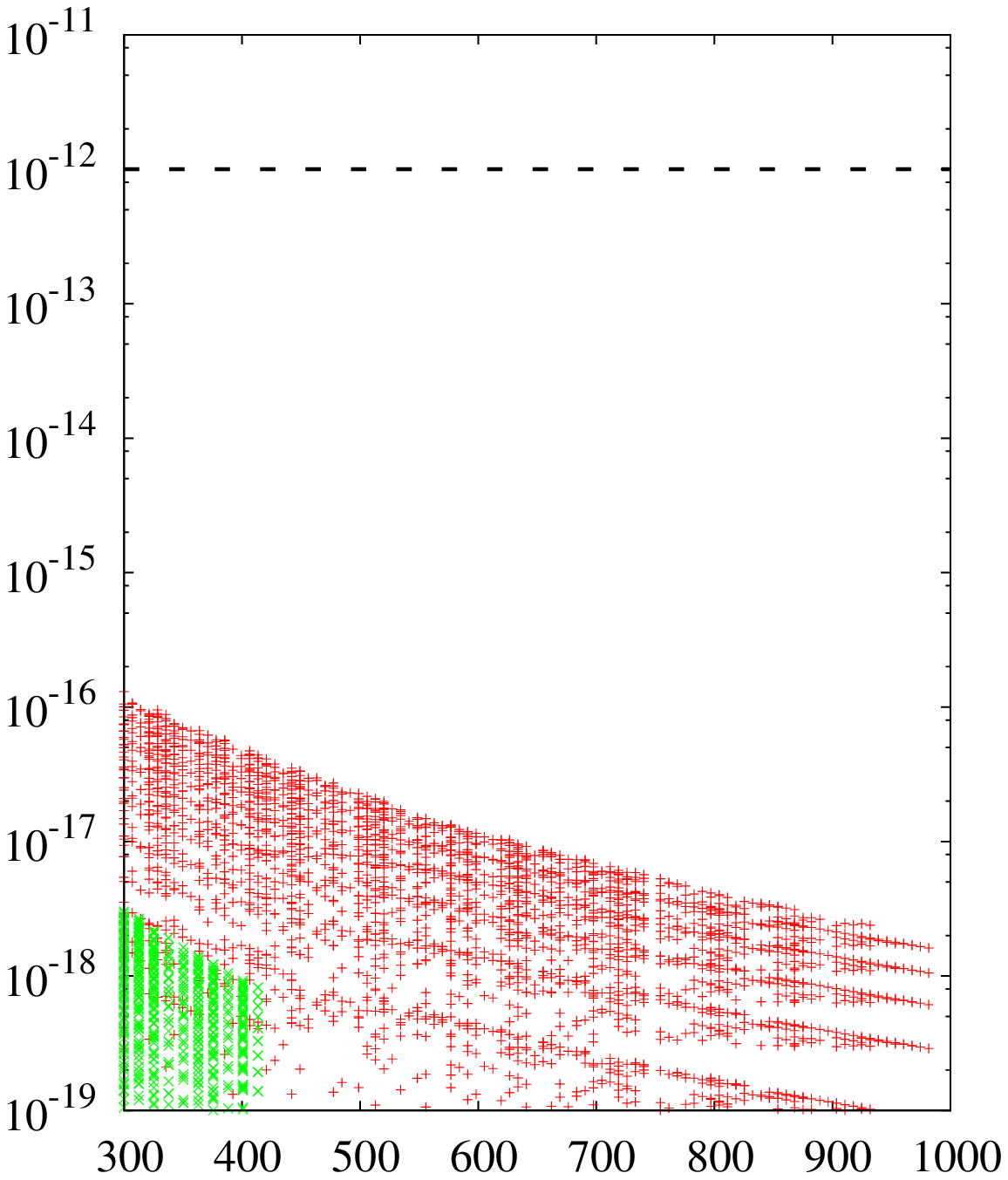,height=3.4in,width=3.2in}}}
\put(210,-300){\makebox(0,0)[bl]{\large{$M_{S_1}$ (GeV)}}}
\put(110,-200){\makebox(0,0)[bl]{\rotatebox{90}{\large{Br$(\mu\to
ee\bar{e})$ }}}}
\put(295,-73){\makebox(0,0)[bl]{\large{ \bf(c)}}}
\put(184,-78){\makebox(0,0)[bl]{\large{Current limit}}}
\put(187,-210){\makebox(0,0)[bl]{{Top-quark}}}
\put(165,-270){\makebox(0,0)[bl]{\rotatebox{90}{{Charm-quark}}}}
\end{picture}
\vspace{11cm} \caption{ Scatter plots of (a) Br($\tau \to
\mu\mu\bar{\mu}$), (b) Br($\tau \to ee\bar{e}$) and (c) Br($\mu
\to ee\bar{e}$) as a function of light leptoquark mass $M_{S_1}$ for
top-quark (red color) and charm-quark (green color) contributions.
The horizontal lines of each plot
 are the current limits of $\tau$ and $\mu$ LFV decay branching ratios.
}
\label{fig-tmg}
%}
\end{figure}
%=========================================================

%=========================================================
\begin{figure}[H] % fig 3
%\FIGURE{
\begin{picture}(320,260)
\put(10,0){\mbox{\psfig{file=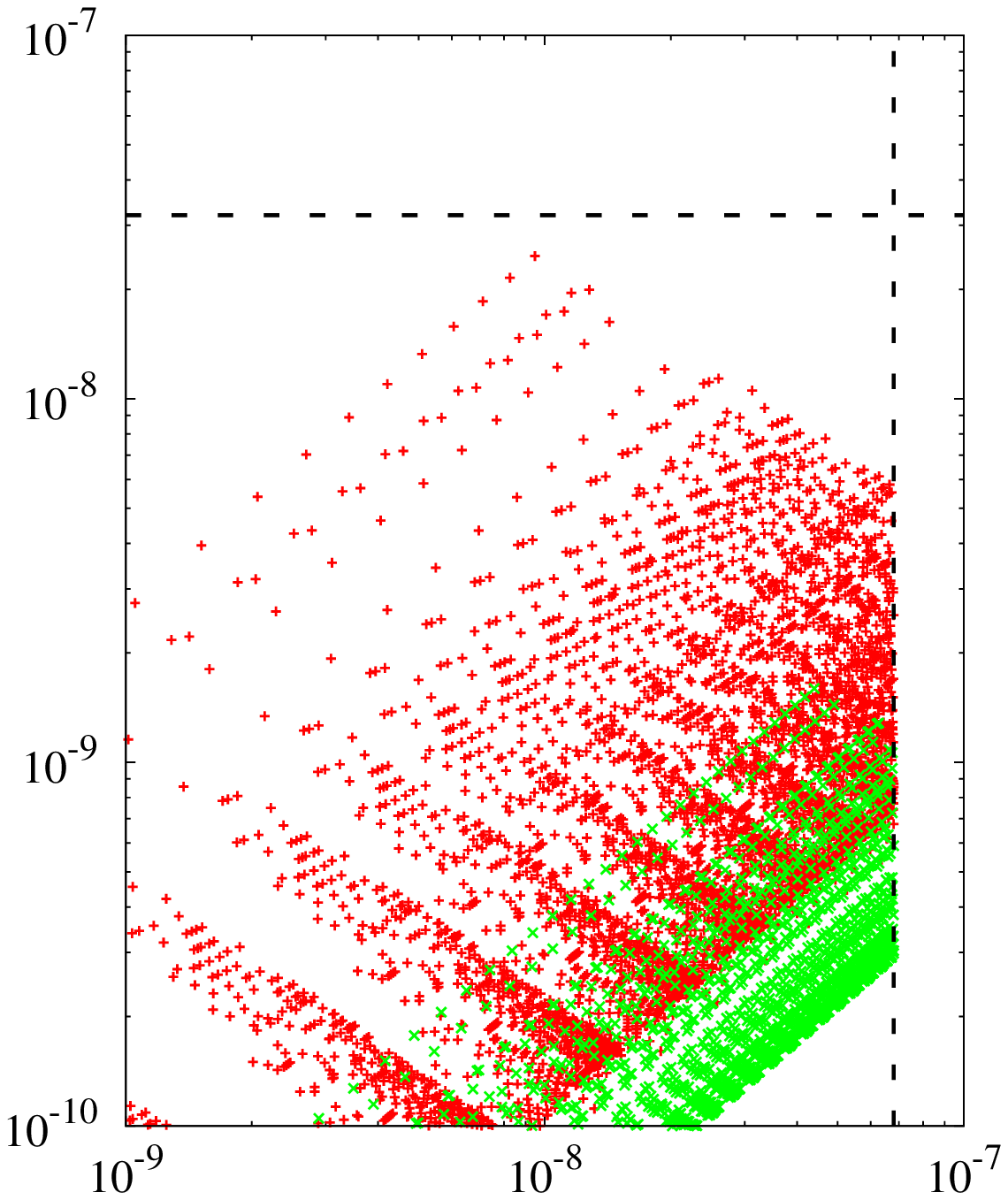,height=3.4in,width=3.2in}}}
\put(165,215){\makebox(0,0)[bl]{\large{\bf(a)}}}
\put(90,-10){\makebox(0,0)[bl]{\large{Br$(\tau \to \mu\gamma)$}}}
\put(57,203){\makebox(0,0)[bl]{\large{Current limit}}}
\put(10,100){\makebox(0,0)[bl]{\rotatebox{90}{\large{Br$(\tau\to
\mu\mu\bar{\mu})$ }}}}
\put(60,130){\makebox(0,0)[bl]{{Top-quark}}}
\put(155,20){\makebox(0,0)[bl]{\rotatebox{45}{{Charm-quark}}}}
\put(240,0){\mbox{\psfig{file=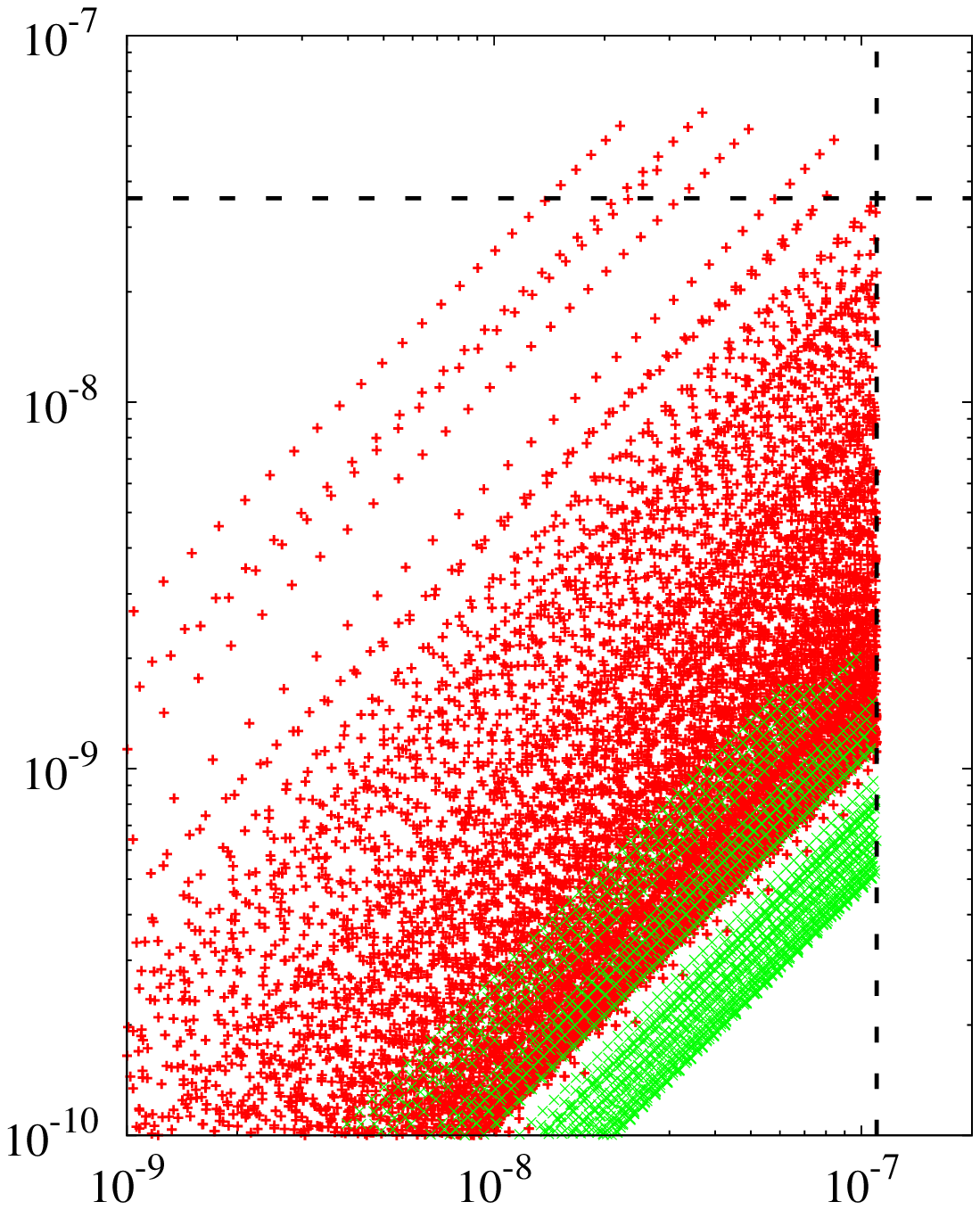,height=3.4in,width=3.2in}}}
\put(343,-10){\makebox(0,0)[bl]{\large{Br$(\tau\to e \gamma)$}}}
\put(390,217){\makebox(0,0)[bl]{\large{\bf(b)}}}
\put(285,206){\makebox(0,0)[bl]{\large{Current limit}}}
\put(290,150){\makebox(0,0)[bl]{{Top-quark}}}
\put(370,20){\makebox(0,0)[bl]{\rotatebox{45}{{Charm-quark}}}}
\put(238,100){\makebox(0,0)[bl]{\rotatebox{90}{\large{Br$(\tau\to
ee\bar{e})$ }}}}
\put(120,-290){\mbox{\psfig{file=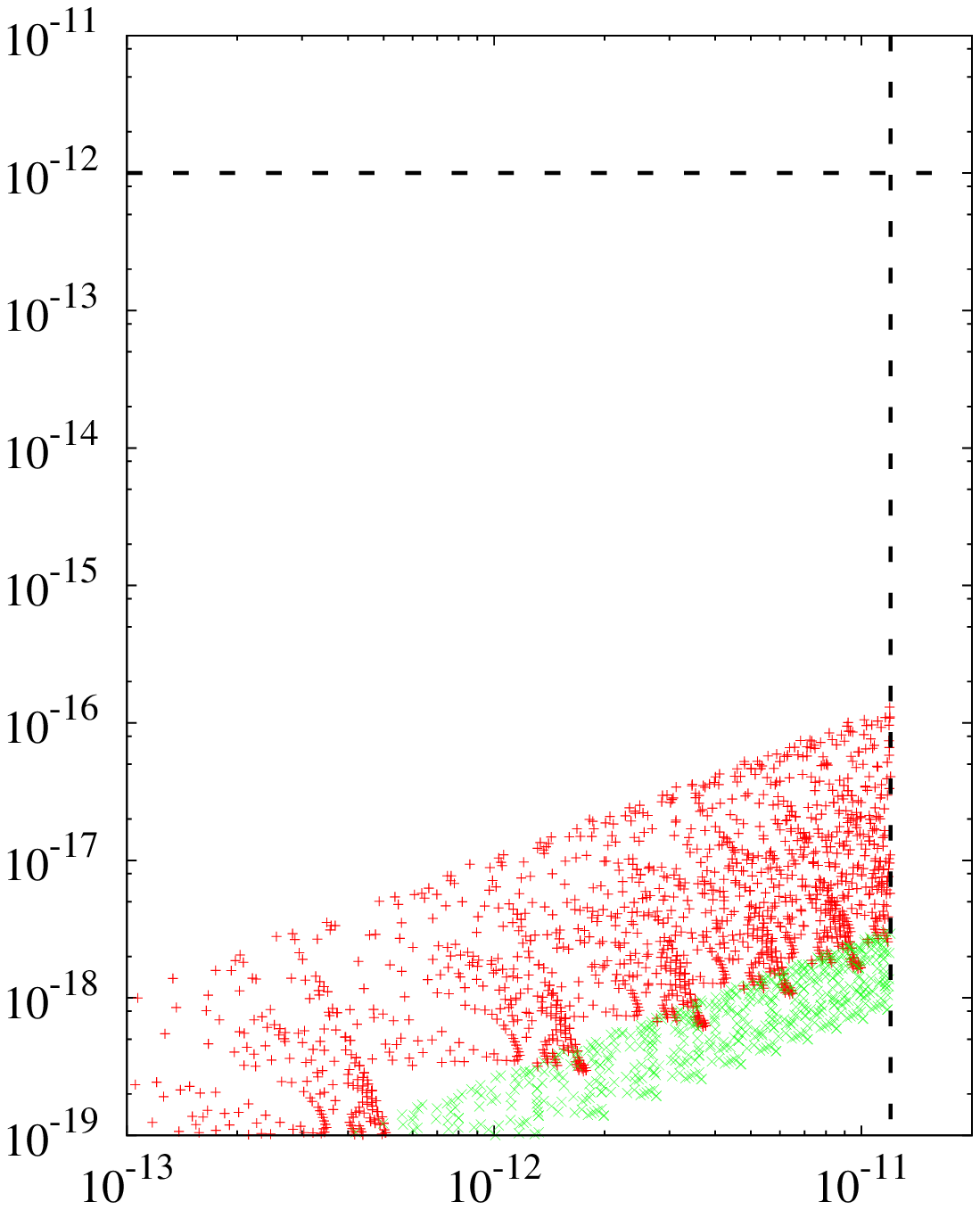,height=3.4in,width=3.2in}}}
\put(203,-300){\makebox(0,0)[bl]{\large{Br$(\mu\to e \gamma)$}}}
\put(106,-200){\makebox(0,0)[bl]{\rotatebox{90}{\large{Br$(\mu\to
ee\bar{e})$ }}}}
\put(165,-80){\makebox(0,0)[bl]{\large{Current limit}}}
\put(265,-73){\makebox(0,0)[bl]{\large{\bf(c)}}}
\put(215,-210){\makebox(0,0)[bl]{{Top-quark}}}
\put(230,-274){\makebox(0,0)[bl]{\rotatebox{23}{{Charm-quark}}}}
\end{picture}
\vspace{11cm} \caption{
Correlations between (a) Br$(\tau\to \mu\mu\bar{\mu})$ and Br$(\tau\to \mu\gamma)$, (b) Br$(\tau\to ee\bar{e})$ and Br$(\tau\to e\gamma)$, (c) Br$(\mu\to ee\bar{e})$ and Br$(\mu\to e\gamma)$. for
top-quark (red color) and charm-quark (green color) contributions
The vertical and horizontal lines correspond to the upper limits
of $\tau$ and $\mu$ LFV decay branching ratios.}
\label{fig-teg}
%}
\end{figure}
%-----------------------------------------------

In Figs~\ref{fig-tmg}, we present our predictions for  the branching
ratios of ($\tau \to \mu\mu\bar{\mu}$) (a) and ($\tau \to
ee\bar{e}$) (b) as a function of light LQ mass $M_{S_1}$ for
top-quark (red color) and charm-quark (green color) contributions. 
These plots have origin in $M_{S_1} = 300$ GeV which roughly corresponds to 
the exclusion limit obtained at HERA~\cite{Wang:2004cj} for leptoquark masses
with couplings of electromagnetic strenght. As we can
see the main contribution comes from the top-quark  contribution 
and can reach 2.47 $\times 10^{-8}$ for {\rm Br}($\tau \to 3\mu$) 
and 5.66 $\times 10^{-8}$ for {\rm Br}($\tau \to 3e$) 
which are comparable with the
present bounds. We find that the main contribution to
 $\tau \to ee\bar{e}$ and $\tau \to \mu\mu\bar{\mu}$ decays is produced from
the photon-penguins diagrams which were not taken into account in
Ref.~\cite{Davidson:1993qk}. In fact, for large LQ mass $(m_q \ll
M_{S_1} )$, the photon-penguins are proportional to
$h^2\log(m_q/M_{S_1})/M^2_{S_1}$ which were known as log
enhancement in the literature~\cite{Kuno:1999jp}. On the other
hand, the naive expectation of Z-penguin and box diagrams leads to
that they are of orders  ${\cal{O}}(h^2 m^2_q/M^4_{S_1})$ and
${\cal{O}}(h^4 m^2_q/M^4_{S_1})$, respectively. 
The same considerations regarding log enhancements hold for the 
$\tau\to e \mu^- \mu^+$ and $\tau\to \mu e^-e^+$ processes. 
However, the uppper limits on the Brs of $\tau\to e \mu^- \mu^+$ and $\tau\to \mu e^-e^+$ could be of ${\cal{O}}(10^{-8})$ and the order in size
is 
{\rm Br}($\tau \to 3e$) $>$ 
{\rm Br}($\tau\to \mu^- e^-e^+$) $>$
{\rm Br}($\tau\to e^- \mu^- \mu^+$) $>$ 
{\rm Br}($\tau \to 3\mu$). Since, $\tau\to e^- e^- \mu^+$ and $\tau\to \mu^- \mu^- e^+$ are induced by box diagrams then they are expected to be small.
On the contrary, since the
current bound on the $\mu \to e \gamma$ decay imposes very strong
constraints on the related couplings, the predicted {\rm Br}($\mu
\to 3e$) is rather too small to be observed.

In Fig.~\ref{fig-teg}, we show the correlations between Br($\tau
\to 3\mu$) and Br($\tau \to \mu\gamma$) in the upper left panel,
(b) Br($\tau \to 3e$) and Br($\tau \to e\gamma$) in the upper
right panel, and (c) Br($\mu \to 3e$) and Br($\mu \to e\gamma$) in
the lower panel. We observe that it is possible to accommodate
both $\tau \to 3\ell$ and $\tau \to \ell\gamma$ branching ratios
for certain choices of LQ parameters. This leads to simple
correlation like
\begin{eqnarray}
\frac{Br(\tau\to3\ell)}{Br(\tau\to \ell\gamma)} \approx {\cal{O}}(10^{-1}), \qquad \frac{Br(\mu\to 3e)}{Br(\mu\to e\gamma)} \approx {\cal{O}}(10^{-3})
\end{eqnarray}
for top-quark contribution, which is in agreement with the ratio expected by the dominance of the Penguin-type Fig.\ref{fig:tau3m-diagrams}(a)-(b).

\section{Conclusion}
We have studied the muon anomalous magnetic moment, lepton flavor
violating muon and tau decays $\ell \to \ell_i \ell_j
\bar{\ell}_j$ and $\ell\to \ell^{\prime} \gamma$ that are generated by scalar LQ
interactions. We have found that scalar LQ can explain the
discrepancy between the experimental value of $(g-2)_\mu$ and  its
standard model prediction without any contradictions  with the
experimental bound of LFV tau decay processes. The present
experimental limits are used to constrain the leptoquark parameter
space. We set equal couplings
%, a choice inspired by
%maximal $\nu_{\mu}-\nu_{\tau}$ mixing, 
and obtain the upper limits
of the different product of leptoquark couplings by confronting
LFV observable with experimental results. 
Our prediction 
is that $\tau\to 3\mu, 3e, e2\mu $ and $\tau\to \mu 2e$ 
get the leading contributions from the so-called photon-penguin diagrams and could be of ${\cal{O}}
(10^{-8})$ which can be accessible by the presents
experiments and the future linear colliders, such as ILC. On the
contrary, the current bounds on LFV impose very strong constraints
on the Br$(\mu\to ee\bar{e})$ and the ratio is too small to be
observed in the near future. Hence any observation of LFV
processes in the charged lepton sector, which are being probed
with ever increasing sensitivity, would unambiguously point to
non-standard interactions. Indeed, such indirect observations taken in isolation may not imply much on the exact nature of new
physics. But a study of possible correlations of its effects on
different independently measured charged LFV observable might
provide a powerful cross-check and lead to identification of new
physics through LHC/LFV synergy.

%============================================================
 \section*{ACKNOWLEDGEMENTS}
We would like to thank the Abdus Salam International Centre for Theoretical
Physics (ICTP) for good hospitality and acknowledge the considerable help of
the High Energy Section. This work was done at the high energy
section within the framework of the associate Scheme.
We would like also thank Chuan-Hung Chen for useful discussions and comments.
R.B. was supported by National Cheng Kung University 
Grant No. HUA 97-03-02-063. 
%R.B and G.F  would like to thank ICTP for the hospitality, 
%where part of this work took place.

%=========================================================

%-----------------------------------------------

%\newpage
\appendix
%=======================================================
\section{Constraint form $\pi \to e\nu_{e}$ and $\pi \to \mu \nu_{\mu}$ decays }
%==================================

We follow \cite{pi_decay,DBC} to constrain leptoquark parameters
using pion decay data. Form the interactions given in
Eq.~(\ref{lag}), we obtain the effective four-Fermi interaction
\begin{eqnarray}
{\mathcal{L}}_{eff} &=& \label{lag2}
-\frac{h'_{ai}h^{'*}_{bj}\Gamma^+_{R,k} \Gamma_{k,R}}{M^2_{S_k}}
(\bar{e}^c_iP_L u_a )(\bar{d}_b P_R \nu^c_j) \\\no &-&
\frac{h_{ai} h^{'*}_{bj}\Gamma^\dagger_{R,k}
\Gamma_{k,L}}{M^2_{S_k}} (\bar{e}^c_iP_R u_a )(\bar{d}_b P_R
\nu^c_j)
\end{eqnarray}
By using the Fierz transformation, we can rewrite Eq.(\ref{lag2})
as
\begin{eqnarray}
{\mathcal{L}}_{eff} &=&\no  -\frac{1}{2 M^2_{S_k}}
h'_{ai}h^{'*}_{bj} \Gamma^\dagger_{R,k} \Gamma_{k,R}(\bar{d}_{L,b}
\gamma_\mu u_{L,a})(\bar{\nu}_{L,j} \gamma^\mu e_{L,i})
\\ &+&\frac{1}{2M^2_{S_k} }h_{ai}h^{'*}_{bj}\Gamma^\dagger_{R,k}
\Gamma_{k,L} (\bar{d}_{L,b}  u_{R,a})(\bar{\nu}_{L,j} e_{R,i})
\end{eqnarray}
On the other hand, the conventional interaction for the $\pi \to
l\nu_{l}$ decay in the SM is given by
\begin{eqnarray*}
{\mathcal{L}}_{eff} &=& -\frac{G_F
V_{ud}}{\sqrt{2}}[\bar{\nu}\gamma_\mu (1-\gamma_5)l]
[\bar{d}\gamma^\mu (1-\gamma_5)u] + {\rm h.c}
\end{eqnarray*}
here $|V_{ud}|$ is the Cabibbo-Kobayashi-Maskawa (CKM) matrix
elements between the constituent of the pion meson and $G_F$ is
the Fermi coupling constant. The ratio $R_{th}$ of the electronic
and muonic decay modes is \cite{Cirigliano:2007xi}
\begin{eqnarray}
R_{th} &=& \no \frac{\Gamma_{SM}(\pi^+ \to
\bar{e}\nu_e)}{\Gamma_{SM}(\pi^+ \to \bar{\mu}\nu_\mu)}
\\\no
&=& \bigg(\frac{m^2_e}{m^2_\mu}\bigg)\bigg(\frac{m^2_\pi -
m^2_e}{m^2_\pi - m^2_\mu}\bigg)^2\bigg(1 + \delta\bigg)
\\
&=& (1.2352 \pm 0.0001) \times 10^{-4}
\end{eqnarray}
where $\delta$ is the radiative corrections, Thus the ratio
$R_{th}$ is very sensitive to non standard model effects (such as
multi-Higges, non-chiral leptoquarks).
The experimental value of the ratio is \cite{pdg}
\begin{eqnarray}
R_{exp} = (1.2302 \pm 0.004) \times 10^{-4}
\end{eqnarray}
The interference between the standard model and LQ model can be
expressed by
%====
%\begin{widetext}
\begin{eqnarray}
R_{SM-LQ} &=& R_{th} +R_{th}\,\frac{m^2_{\pi^+}}{m_u + m_d}
\bigg(\frac{1}{\sqrt{2}}\frac{{\rm Re} (h_{ue}h^{'*}_{ue})}{G_F
V_{ud}M^2_{S_k}} \frac{1}{m_{e}}-\frac{1}{\sqrt{2}} \frac{{\rm Re}
(h_{u\mu}h^{'*}_{u\mu})}{G_F
V_{ud}M^2_{S_k}}\frac{1}{m_{\mu}}\bigg)
\Gamma^\dagger_{R,k}\Gamma_{k,L}\nonumber\\
\end{eqnarray}
%====
At 2$\sigma$ level, we get
\begin{eqnarray}
 R_{min} < \sum^{2}_{k=1}\bigg(\frac{m_{\pi}}{m_{e}} \frac{{\rm Re}
 (h_{ue}h^{'*}_{ue})}{M^2_{S_k}}-\frac{m_{\pi}}{m_{\mu}} \frac{{\rm Re} (h_{u\mu}h^{'*}_{u\mu})}{ M^2_{S_k}}\bigg)
 \Gamma^\dagger_{R,k}\Gamma_{k,L} < R_{max}
\end{eqnarray}
%\end{widetext}
where,
\begin{eqnarray}
R_{min} &=& -1.06 \times 10^{-8} {\rm GeV}^{-2},\\  R_{max}  &=&
2.45 \times 10^{-9} {\rm GeV}^{-2}.
\end{eqnarray}
The total contribution to $R_{SM-LQ}$ must be smaller than the
differences between SM and experiment within the allowed error
limits.

%=======================================================
%================================
\section{One loop functions}
%--------------------------------------------
The loop functions used in text are given by
\begin{eqnarray}
F_{1}(x) &=& \frac{\big[2 + 3x -6x^2+x^3+6x\log(x)\big]}{12(1-x)^4},\\
F_{2}(x) &=& \frac{\big[1 - 6x +3x^2+2x^3-6x^2\log(x)\big]}{12(1-x)^4},\\
F_{3}(x) &=& \frac{-1}{2(1-x)^3}\big[ 3 -4x + x^2 + 2 \log(x)\big],\\
F_{4}(x) &=& \frac{1}{2(1-x)^3}\big[ 1 - x^2 + 2 x \log(x)\big],\\
F_5(x) &=& \frac{1}{36(x-1)^4}\big[ 16 - 45 x + 36 x^2 - 7 x^3 + 6
(2 - 3 x) \log(x)\big]\\
F_6(x) &=& \frac{\big[ -2 + 9 x -18 x^2 + 11 x^3 - 6 x^3 \log(x)
\big]}{36(x-1)^4} \\
F_7 (x) &=& \frac{1-x+\log(x)}{(x-1)^2}\\
F_8 (x) &=& \frac{1}{8(x-1)^2}\bigg[ 3 - 4x + x^2 +4x\log(x)-2x^2\log(x)\bigg]
\end{eqnarray}
%D_0 (m_1^2,m_2^2,m_3^2,m_4^2) &=& -\frac{m_1^2\log(m_1^2)}{(m_1^2-m_2^2)(m_1^2-m_3^2)(m_1^2-m_4^2)} - %\frac{m_2^2\log(m_2^2)}{(m_2^2-m_1^2)(m_2^2-m_3^2)(m_2^2-m_4^2)}\nonumber\nonumber\\&-& %\frac{m_3^2\log(m_3^2)}{(m_3^2-m_1^2)(m_3^2-m_2^2)(m_3^2-m_4^2)}
%-\frac{m_4^2\log(m_4^2)}{(m_4^2-m_1^2)(m_4^2-m_2^2)(m_4^2-m_3^2)}\nonumber\\
\begin{eqnarray}
D_0 (x,y,z,k) &=& -\frac{x \log (x)}{(x-k) (x-y) (x-z)}-\frac{y
\log (y)}{(y-k) (y-x) (y-z)}\nonumber\\&-&\frac{z \log (z)}{(z-k)
(z-x) (z-y)}-\frac{k \log (k)}{(k-x) (k-y) (k-z)}\\
%----------------
 D_0 (y,y,z,k)
&=& \frac{1}{(k-y)^2 (k-z) (y-z)^2}\bigg[-k \log (k)
(y-z)^2+\big(-z k^2 \nonumber \\ &+& \left(y^2+z^2\right) k-y^2 z\big) \log
(y)\nonumber \\ &+&(k-y) ((k-z) (y-z)+(k-y) z \log
   (z))\bigg]\\
%--------------
D_0 (x,y,k,k) &=&\frac{1}{(k-x)^2 (k-y)^2 (x-y)}\big[-x \log (x) (k-y)^2\nonumber\\
&+&(x-y) \left(k^2-x y\right) \log (k)-(k-x) ((k-y) (x-y)\nonumber \\ &+& (x-k) y \log
   (y))\big]\\
 \widetilde{D}_0 (x,y,z,k) &=& -\frac{x^2 \log (x)}{(x-k) (x-y) (x-z)}-\frac{y^2
 \log (y)}{(y-k) (y-x) (y-z)}\nonumber\\&-&\frac{z^2 \log (z)}{(z-k) (z-x)
 (z-y)}-\frac{\log (k) k^2}{(k-x) (k-y) (k-z)}\\
%---------------
 \widetilde{D}_0 (y,y,z,k) &=&\frac{1}{(k-y)^2 (k-z)(y-z)^2}\big[-k^2 \log (k) (y-z)^2
 \nonumber \\ &+& y (k-z) (k (y-2 z)+y z) \log (y) \nonumber \\&+& (k-y) \left((k-y) \log (z)
  z^2+y (k-z) (y-z)\right)\big]\\
%\end{eqnarray}
%
%  \begin{eqnarray}
 \widetilde{D}_0 (x,y,k,k) &=&\frac{1}{(k-x)^2 (k-y)^2(x-y)}\big[-x^2 \log (x) (k-y)^2 \nonumber \\ &+& k (x-y) (k (x+y)-2 x y) \log (k)\nonumber\\&-&(k-x) \left((x-k) \log (y) y^2+k (k-y) (x-y)\right)\big]
%\tilde{D}_0 (m_1^2,m_2^2,m_3^2,m_4^2) &=& -\frac{m_1^4\log(m_1^2)}{(m_1^2-m_2^2)(m_1^2-m_3^2)(m_1^2-m_4^2)} - %\frac{m_2^4\log(m_2^2)}{(m_2^2-m_1^2)(m_2^2-m_3^2)(m_2^2-m_4^2)}\nonumber\\&-& %\frac{m_3^4\log(m_3^2)}{(m_3^2-m_1^2)(m_3^2-m_2^2)(m_3^2-m_4^2)}
%-\frac{m_4^4\log(m_4^2)}{(m_4^2-m_1^2)(m_4^2-m_2^2)(m_4^2-m_3^2)},\nonumber\\
\end{eqnarray}
%======================================================
%=======================================================

%=================================================
\end{document}